 \documentclass[final,5p,times,twocolumn]{elsarticle}

\usepackage{amssymb}
\usepackage{amsmath}
\usepackage{hyperref}
\usepackage{graphicx}
\usepackage{rotating}

\usepackage{tabularx, booktabs}
\usepackage{array}
\usepackage{longtable}   
\usepackage{makecell}
\usepackage{ragged2e}
\usepackage{changepage}
\usepackage{tabularx}
\usepackage{eurosym}
\usepackage{multirow} 
\usepackage{color, soul}

\usepackage{threeparttable}
\usepackage{pifont}
\usepackage{makecell}
\usepackage{tikz}  
\usepackage{comment}

\newcommand{\cmark}{\ding{51}}     
\newcommand{\nmark}{\textendash} 
\usepackage{tikz}
\newsavebox{\pmarkbox}
\savebox{\pmarkbox}{%
  \tikz[baseline=(c.base)]{%
    \node[draw, circle, inner sep=0pt, minimum size=1.1ex,
          path picture={%
            \fill (path picture bounding box.north)
                  arc (90:270:0.55ex) -- cycle;
          }] (c) {};%
  }%
}
\newcommand{\pmark}{\usebox{\pmarkbox}}

\journal{Renewable and Sustainable Energy Reviews}

\begin{document}

\begin{frontmatter}



\title{Multidimensional Resilience for Electrical Power Systems: Systematic Review, Integrated Index, and Validation under Real-World Cyber-Physical Attack Scenarios}

\author{Isaac Ortega Romero}
\ead{I.OrtegaRomero001@umb.edu}

\author{Ioannis Zografopoulos}
\ead{I.Zografopoulos@umb.edu}

\affiliation{
    organization={Engineering Department, University of Massachusetts Boston},
    addressline={100 William T. Morrissey Blvd},
    city={Boston},
    state={MA},
    postcode={02125},
    country={USA}
}

\begin{abstract}
The accelerating decarbonization of energy systems has transformed electrical power systems into complex infrastructures exposed to threats whose interactions generate systemic vulnerabilities that conventional resilience approaches fail to capture. Although resilience assessment has expanded across multiple dimensions, existing studies largely examine them in isolation or adjacent pairs, leaving cross-dimensional couplings insufficiently explored. 
This study demonstrates \emph{i)} that single-dimension assessments fail to capture the degradation produced by simultaneous cross-dimensional failures, \emph{ii)} the nonlinear amplification emerging when physical, operational, and digital-cyber dimensions are jointly compromised, and \emph{iii)} the intensification imposed by climatic and economic-regulatory stressors.

To this end, we leverage a hybrid quantitative methodology. A PRISMA 2020 review with backward and forward snowballing identifies methodological gaps and unresolved dependencies across five resilience dimensions: \textit{physical, operational, digital-cyber, climatic-external,} and \textit{economic-regulatory}. Following this analysis, a Multidimensional Resilience Index ($\mathcal{MDRI}$) is developed to capture endogenous couplings and exogenous amplification effects and is validated under escalating cyber-physical attack scenarios inspired by the December 2025 attack on Polish energy infrastructure. Results show that degradation under cascading and simultaneous failures is nearly eight times greater than under isolated stress, while exogenous conditions amplify degradation by an additional factor approaching six, with 72\% of this amplification driven by exogenous stressors. Combined, these mechanisms produce a 46-fold increase in resilience loss compared to a single-vector reference. The proposed framework advances resilience assessment by providing an integrated, quantitative basis for identifying compound vulnerabilities, evaluating cross-dimensional amplification mechanisms, and reframing resilience from a discipline-specific attribute into a \textit{cross-dimensional} property of modern power systems.\\


\end{abstract}


\end{frontmatter}



\section{Introduction} \label{s:Intro}

The transition toward decarbonized energy systems has fundamentally redefined the role of the electrical grid in modern society. The accelerated electrification of sectors historically dependent on fossil fuels, combined with the massive integration of variable and distributed generation, has transformed the grid from a passive energy delivery infrastructure into a structural backbone of contemporary economic, industrial, and digital activity~\cite{XU2021111642, MOHSENI2022112095, VALLEJODIAZ2024114525}.
This transformation occurs amid growing operational complexity, driven by the progressive digitalization of the system and its deepening interdependencies with other critical infrastructures, which amplify the potential consequences of any supply disruption~\cite{goteman2025offshore, cavus2026digital}. As perturbations of diverse nature, climatic, physical, and cyber-related, increasingly converge, approaches centered exclusively on reliability 
prove insufficient. Thus, resilience, i.e.,   the system's capacity to absorb disturbances, maintain essential functions, and restore operation in a timely and adaptive manner, emerges as a fundamental criterion~\cite{paul2024resilience, monie2025generic}. \looseness=-1

\begin{table*}[t]
\centering
\caption{Coverage of resilience dimensions across resilience-focused studies published in \textit{Renewable and Sustainable Energy Reviews} 
(2021--2026).}

\label{tab:comparison}
\footnotesize
\setlength{\tabcolsep}{4pt}
\renewcommand{\arraystretch}{1.0}
\begin{threeparttable}
\begin{tabular}{@{}ccccccccc@{}}
\toprule
  \makecell{\textbf{Physical}} 
  & \makecell{\textbf{Operational}} 
  & \makecell{\textbf{Digital-}\\\textbf{Cyber}} 
  & \makecell{\textbf{Climatic-}\\\textbf{External}} 
  & \makecell{\textbf{Economic-}\\\textbf{Regulatory}} 
  & \makecell{\textbf{Cross-}\\\textbf{dim.}} 
  & \makecell{\textbf{Multi-dim.}\\\textbf{index}} 
  & \makecell{\textbf{Case}\\\textbf{study}}
  & \textbf{Reference} \\
\midrule
\cmark & \nmark & \pmark & \cmark & \nmark & \nmark & \nmark & \nmark & 
~\cite{jasiunas2021energy} \\
\cmark & \pmark & \nmark & \pmark & \nmark & \nmark & \cmark & \nmark & 
~\cite{ahmadi2021frameworks} \\
\cmark & \cmark & \cmark & \pmark & \nmark & \cmark & \nmark & \nmark & 
~\cite{XU2021111642} \\
\cmark & \cmark & \nmark & \cmark & \nmark & \nmark & \pmark & \nmark & 
~\cite{wang2022systematic} \\
\pmark & \nmark & \cmark & \nmark & \nmark & \nmark & \nmark & \nmark & 
~\cite{hou2024cyber} \\
\cmark & \cmark & \nmark & \cmark & \nmark & \nmark & \nmark & \nmark & 
~\cite{tang2024fault} \\
\cmark & \pmark & \nmark & \cmark & \nmark & \nmark & \pmark & \nmark & 
~\cite{paul2024resilience} \\
\pmark & \cmark & \nmark & \nmark & \nmark & \nmark & \nmark & \nmark & 
~\cite{cao2025secure} \\
\cmark & \cmark & \nmark & \cmark & \nmark & \nmark & \nmark & \nmark & 
~\cite{zidane2025microgrids} \\
\cmark & \pmark & \nmark & \cmark & \nmark & \nmark & \nmark & \nmark & 
~\cite{goteman2025offshore} \\
\cmark & \pmark & \nmark & \cmark & \nmark & \pmark & \cmark & \nmark & 
~\cite{monie2025generic} \\
\cmark & \cmark & \pmark & \cmark & \nmark & \pmark & \nmark & \pmark & 
~\cite{cavus2026digital} \\
\midrule
\textbf{\cmark} & \textbf{\cmark} & \textbf{\cmark} 
 & \textbf{\cmark} & \textbf{\cmark} 
 & \textbf{\cmark} & \textbf{\cmark} & \textbf{\cmark} & \textbf{This work} \\
\bottomrule

\end{tabular}
\begin{tablenotes}
\footnotesize
\item \cmark~fully addressed\quad \quad \quad \pmark~partially addressed\quad \quad \quad \nmark~not addressed.
\end{tablenotes}
\end{threeparttable}
\end{table*}

Building on this understanding, a review of existing resilience studies reveals five recurrent analytical dimensions~\cite{zidane2025microgrids, cavus2026digital}. 
However, a substantial portion of this literature has addressed these dimensions in a fragmented manner, developing metrics, models, and frameworks specific to each perspective, overlooking cross-domain interdependencies. This separation hinders the understanding of coupled phenomena, particularly cascading failures, in which an initial disturbance propagates across infrastructures, operational layers, and control systems, progressively amplifying its effects on electrical power systems (EPS)~\cite{tang2024fault, cao2025secure}. Consequently, a system considered resilient under dimension-specific criteria may exhibit critical vulnerabilities once inter-domain interactions are accounted for.

This methodological gap highlights the need for integrated approaches that simultaneously represent the distinct resilience dimensions of EPS, model their interdependencies, and translate their combined effects into coherent quantitative measures. Existing reviews have examined specific subsets of this problem in depth, primarily physical-climatic, physical-operational, and cyber-physical interactions. However,  as demonstrated in Table~\ref{tab:comparison}, the \emph{cross-dimensional integration} of all relevant dimensions and the representation of amplification mechanisms under compound disturbances remain systematically unaddressed in the literature~\cite{tang2024fault, cao2025secure}.

The present study develops a systematic review that consolidates the state of the art on EPS' resilience while identifying interaction patterns, methodological fragmentation, and unresolved cross-dimensional dependencies within existing resilience assessments. The reviewed literature is organized into five analytical dimensions: \emph{i)} physical, \emph{ii)} operational, \emph{iii)} digital-cyber, \emph{iv)} climatic-external, and \emph{v)} economic-regulatory. Beyond identifying these gaps, the synthesis reveals the absence of generalized formulations capable of representing disturbance propagation across resilience dimensions quantitatively. Therefore, this study proposes a quantitative framework to model inter-dimensional couplings and assess their contribution to the systemic degradation dynamics of EPS. Namely, the contributions of this work are as follows:

\begin{itemize}

\item \textbf{Multidimensional systematic analysis:} Review of 120 studies selected through the PRISMA 2020 protocol and hybrid snowballing method, organized into five dimensions, with systematic identification of quantitative metrics, modeling approaches, inter-dimensional dependencies, and recurring methodological limitations across each dimension.

\item \textbf{Integrated coupling framework and multidimensional resilience index:} Development of a quantitative formulation representing unidirectional and bidirectional interactions among endogenous and exogenous dimensions, from which a comprehensive multidimensional resilience index ($\mathcal{MDRI}$) is derived, incorporating cross-dimensional amplification effects and ensuring comparable assessments across varying coupling regimes.

\item \textbf{Illustrative evaluation under compound stress scenarios.} Evaluation of the proposed framework on a modified IEEE 39-bus system under three escalating cyber-physical attack scenarios, demonstrating how cross-dimensional coupling amplifies systemic degradation in ways that conventional single-dimension assessments cannot capture.

\end{itemize}

\begin{figure*}[t]
\centerline{\includegraphics[width=1\linewidth]{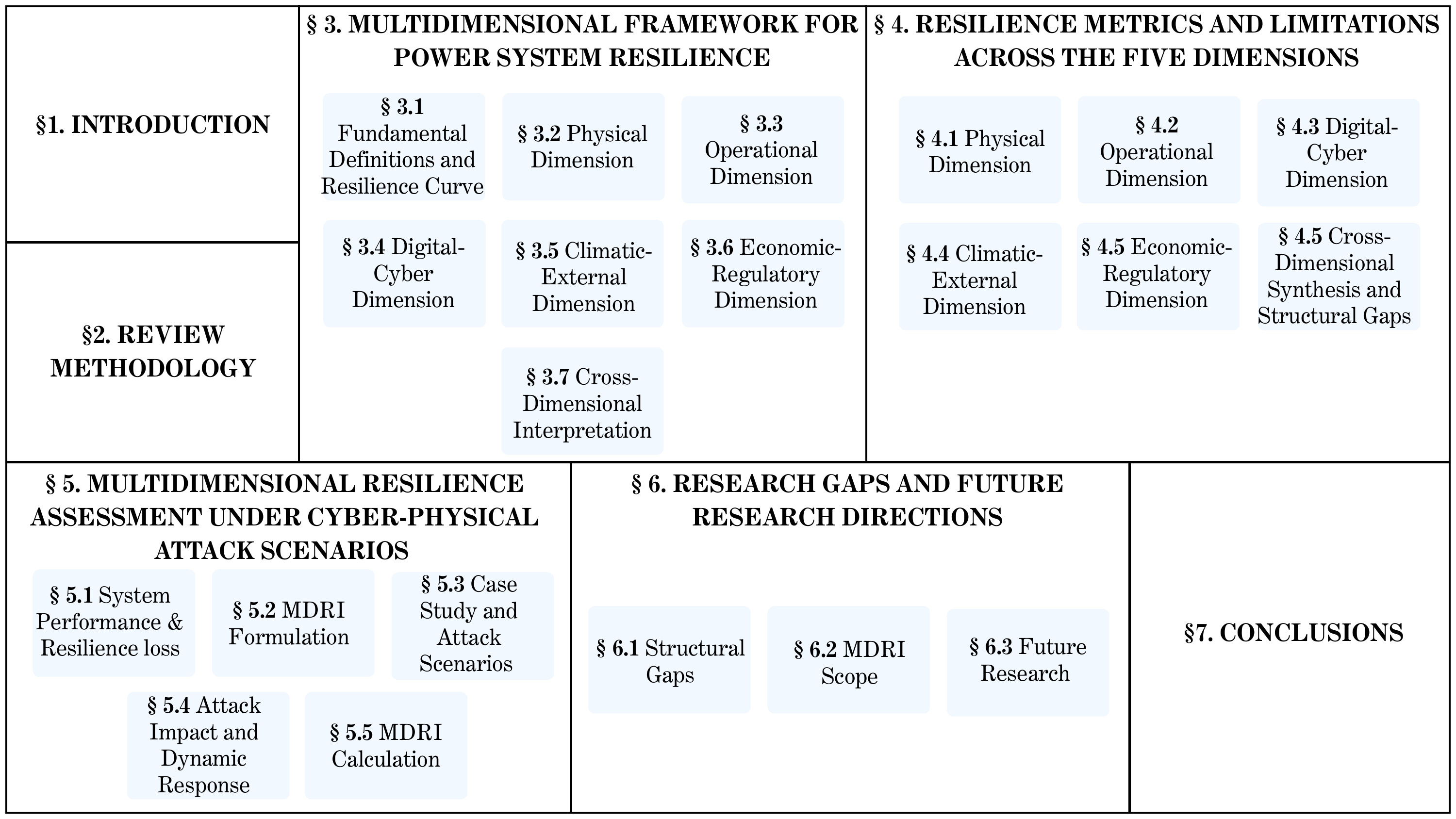}}
\caption{Roadmap of this work.}
\label{fig:roadmap}
\end{figure*}

A schematic overview of this paper is illustrated in Figure~\ref{fig:roadmap}. The rest of the paper is organized as follows: Section \ref{s:Meth} presents the review methodology, Section \ref{s:Mult} introduces the multidimensional framework, Section \ref{s:syst} reviews resilience metrics and limitations across the five dimensions, Section \ref{s:Comp} develops the $\mathcal{MDRI}$ and applies it to cyber-physical attack case studies, Section \ref{s:Futur} outlines research gaps and future research directions, and Section \ref{s:Con} concludes the paper.

\section{Review Methodology} \label{s:Meth}

This review study adopts a systematic methodology to identify, screen, classify, and synthesize the literature on resilience in EPS, following the \textit{PRISMA 2020} guidelines and backward and forward snowballing method to ensure transparency and reproducibility~\cite{Page2021PRISMA}. Rather than performing a quantitative meta-analysis, the objective is to examine how resilience has been conceptualized, modeled, and evaluated across five dimensions identified recurrently in the literature.
The review is guided by four research questions (RQ):
\begin{itemize}
\item \textbf{RQ1:} How is resilience in EPS defined and operationalized in the current literature?
\item \textbf{RQ2:} Which resilience dimensions and cross-dimensional interactions are predominantly addressed?
\item \textbf{RQ3:} What metrics, methods, and analytical approaches are used to assess resilience under compound disruptions?
\item \textbf{RQ4:} To what extent do existing studies enable the formulation of unified quantitative representations of multidimensional resilience interactions?
\end{itemize}

\begin{figure}[t]
\centerline{\includegraphics[width=1\linewidth, trim={0pt 60pt 0pt 30pt}, clip]{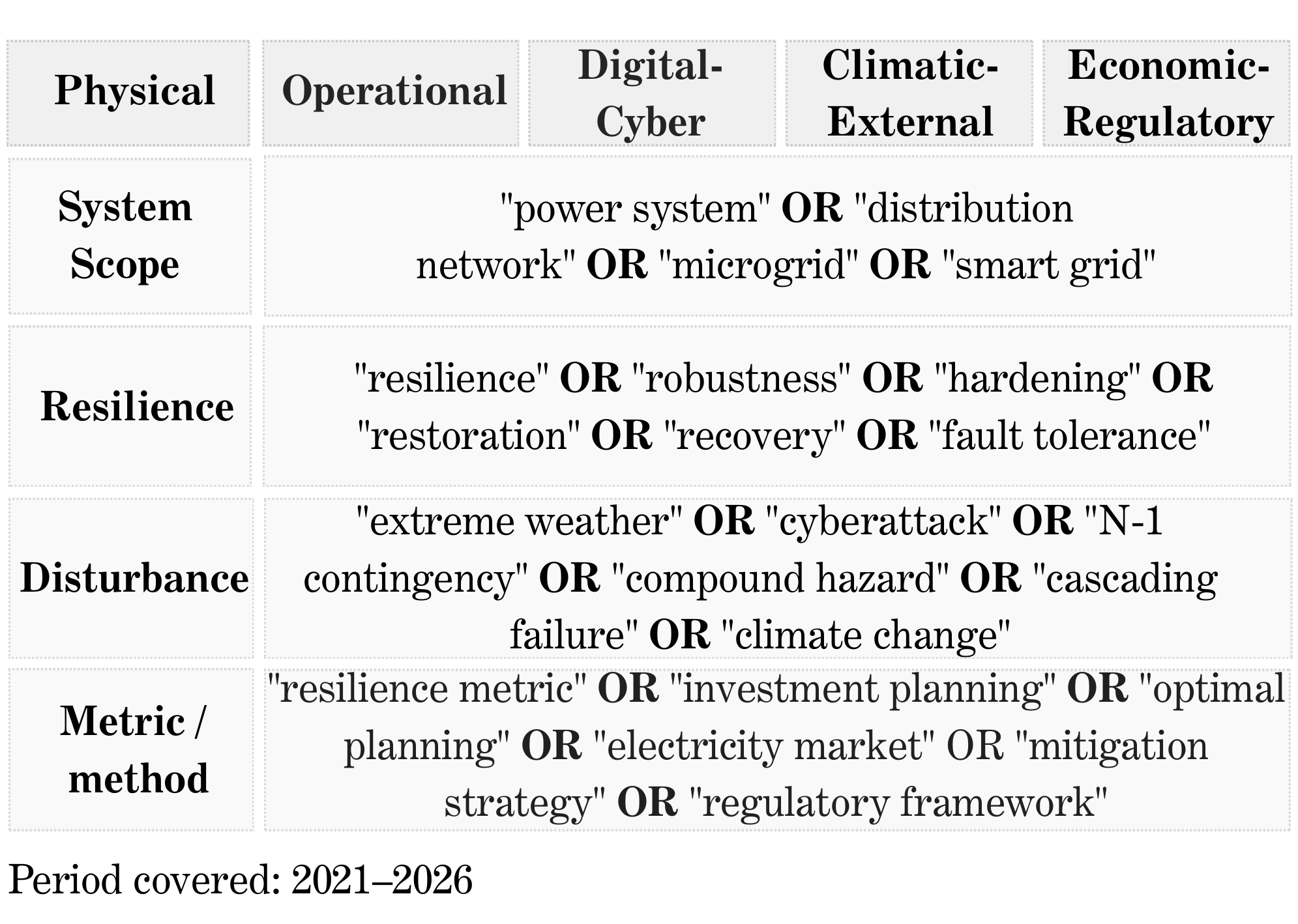}}
\caption{Search strategy for the review of EPS' resilience literature.}
\label{fig:search_strategy}
\end{figure}

The literature search was conducted across four scientific databases, Web of Science, Scopus, IEEE Xplore, and ScienceDirect, supplemented by targeted Google Scholar queries for cross-validation purposes, yielding 18,460 initial records. The search covers the period 2021--2026, extended through backward snowballing to retrieve relevant foundational studies ~\cite{Wohlin2014Snowballing}. 
Search queries were organized around four thematic groups: system scope, resilience concepts, disturbance types, and assessment methods, as shown in Figure~\ref{fig:search_strategy}.

\begin{figure}[t]
\centerline{\includegraphics[width=1\columnwidth]{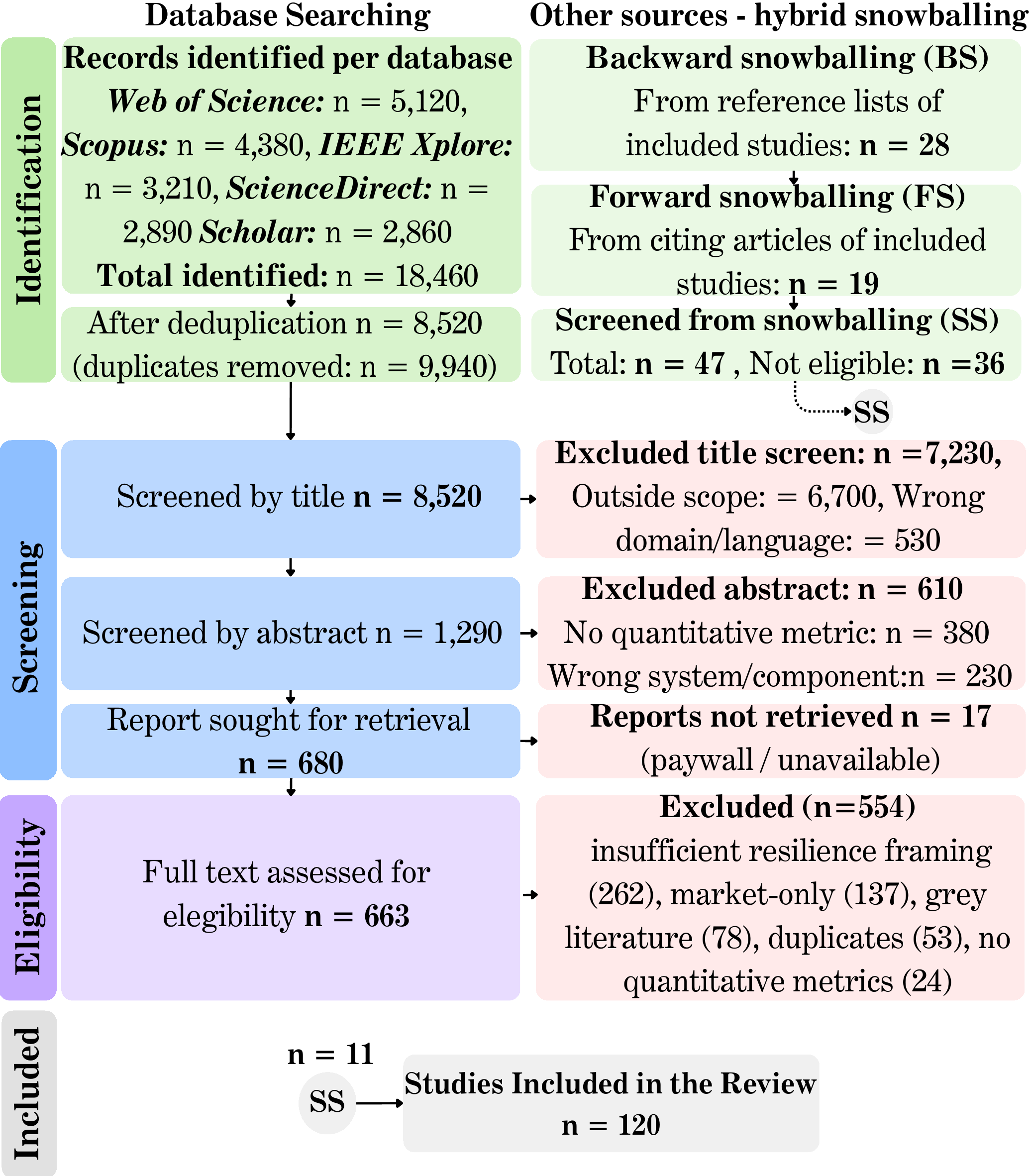}}
\caption{PRISMA 2020 screening flow for the review of EPS' resilience literature~\cite{Page2021PRISMA,Wohlin2014Snowballing}.}
\label{fig:prisma}
\end{figure}

Studies were included if they explicitly addressed at least one resilience dimension of EPS and proposed or evaluated resilience metrics or frameworks. Records were excluded if they \emph{i)}~focused on isolated technical problems without an explicit resilience framing, \emph{ii)}~addressed market analyses unrelated to resilience considerations, \emph{iii)}~consisted of grey literature or duplicate records, or \emph{iv)}~lacked explicit assessment criteria or quantitative metrics. After duplicate removal and eligibility screening, 120 studies were retained, as summarized in Figure~\ref{fig:prisma}.

Each retained study was classified 
according to six attributes: dimensional scope, disturbance type, modeling formulation, resilience metrics, mitigation strategies, and methodological limitations. This characterization forms the basis of the taxonomies developed in Section~\ref{s:syst} to address RQ1--RQ4.

\section{Multidimensional Framework for Power System Resilience}
\label{s:Mult}

\subsection{Fundamental Definitions and Resilience Curve}
Resilience in EPS is defined as ``\emph{the ability of the system and its components to anticipate, withstand, absorb, and respond to extreme events, preserving essential functions with minimal performance degradation, even under severe conditions}" \cite{CIGRE_C4.47_2019}. Resilience also encompasses adaptation, learning, and transformation processes aimed at reducing vulnerability to future events \cite{DWIVEDI2025125001, MACMILLAN2022112841, GOTEMAN2025115649}.
From a temporal perspective, resilience is analyzed through the \textit{resilience curve}, which describes system performance $\Phi(t)$ before, during, and after a disturbance, capturing loss of functionality, absorption capacity, recovery speed, and post-restoration performance \cite{DWIVEDI2025125001, GOTEMAN2025115649}.
The resilience of EPS cannot be assessed as a static attribute 
but as a dynamic and time-dependent process that unfolds in response to disruptive events. This temporal nature is increasingly relevant in modern EPS, characterized by higher interconnection, growing penetration of power electronics and renewable energy resources, increasing digitalization, and growing exposure to extreme events. Under such conditions, traditional resilience approaches -- based on physical models and static operational assumptions -- struggle to represent the evolving system behavior \cite{Xu2025Theoretical}. \looseness=-1

Recent literature has evolved from static interpretations toward the concept of dynamic resilience. Nonetheless, a considerable number of existing studies continue to address it in a fragmented manner \cite{Xu2025Theoretical,JASIUNAS2021111476}. This fragmentation is problematic given the heterogeneous nature of EPS,  comprising generation units, networks, and loads with distinct characteristics and vulnerabilities, meaning system response depends not only on the type of disruptive event but also on the specific components affected \cite{WANG2022112567}.
Current approaches do not adequately capture interactions among these elements nor how multiple dimensions simultaneously influence system response.

\begin{figure}[t]
    \centering
    \includegraphics[width=1\linewidth, trim=0 0 0 0 0, clip]{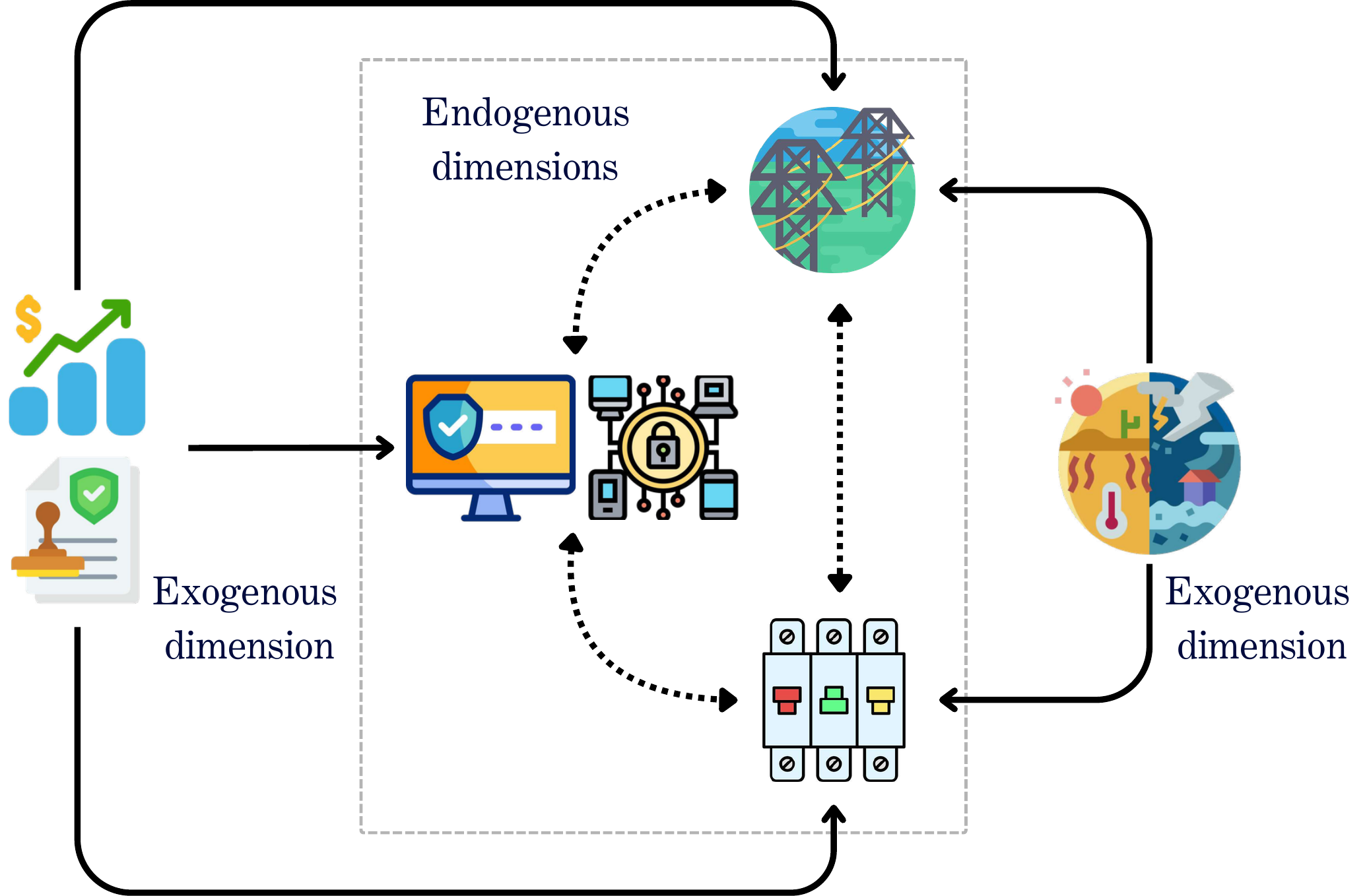}
    \caption{Unidirectional and bidirectional interactions among EPS dimensions.}
    \label{fig:2}
\end{figure}

These limitations underscore the necessity of a multidimensional and dynamic resilience framework structured around \textit{five dimensions}: \emph{i)} physical, \emph{ii)} operational, \emph{iii)} digital-cyber, \emph{iv)} climatic-external, and \emph{v)} economic-regulatory, analyzed in an integrated manner because they define how disruptions affect system performance $\Phi(t)$, how the system responds during the event, and how it adapts afterward. Figure~\ref{fig:2} conceptualizes EPS' resilience as the result of asymmetric and non-uniform interactions across these dimensions, grouped into two categories: \emph{i)} bidirectional couplings among the physical, operational, and digital-cyber dimensions, understood as \textbf{endogenous dimensions}, through which vulnerabilities are dynamically co-produced and \emph{ii)} unidirectional influences through which climatic-external and economic-regulatory factors,  understood as \textbf{exogenous dimensions}, impose constraints on system behavior. 

Within this conceptual structure, the bidirectional couplings are characterized as follows:

\begin{itemize}
    \item \textbf{Physical $\leftrightarrow$ Operational:} The condition of physical assets constrains operational flexibility, network reconfiguration options, and restoration feasibility. Conversely, operational decisions, such as switching actions or load transfers, directly influence equipment loading and stress levels, potentially accelerating physical degradation.

    \item \textbf{Operational $\leftrightarrow$ Digital-Cyber:} Operational constraints, such as limited visibility or damaged infrastructure, may disrupt or isolate cyber systems. In contrast, limitations or failures in digital-cyber systems can directly constrain operational decision-making, coordination, and responses.

    \item \textbf{Digital-Cyber $\leftrightarrow$ Physical:} Cyberattacks or control system failures may propagate into the physical layer through incorrect actuation or delayed responses. Conversely, physical degradation can impair sensor accuracy, communication reliability, and data availability, thereby reducing situational awareness.
\end{itemize}

The climatic-external and economic-regulatory dimensions act as exogenous drivers. Climatic conditions affect physical integrity and operational performance through extreme events and long-term trends, while regulatory frameworks and investment mechanisms impose structural constraints on infrastructure enhancement, digital modernization, and long-term adaptation, limiting the available action space for improving resilience.

\subsection{Physical dimension}
\paragraph{Definition and Importance} The physical dimension refers to the material and structural integrity of assets that enable the generation, transmission, distribution, storage, and supply of electricity throughout energy systems \cite{JASIUNAS2021111476}. It determines the system's initial exposure and susceptibility to disturbances, shaping the extent of immediate impacts on infrastructure and service continuity.

\paragraph{Limitations} The physical dimension does not capture how the system coordinates, adapts, or recovers once damage has occurred. Approaches focused exclusively on infrastructure hardening assume that physical robustness alone ensures resilience, overlooking dynamic interactions with operational decision-making, digital control systems, and external constraints. As a result, purely physical assessments may underestimate cascading failures or overestimate recovery capacity \cite{JASIUNAS2021111476}.

\paragraph{Real World Implications} The 2010 Chile earthquake illustrates this weakness. The event caused extensive structural damage to substations, transmission lines, and distribution infrastructure, resulting in widespread service interruptions. Although the infrastructure withstood significant damage, the system lacked operational flexibility and coordinated recovery mechanisms, preventing a timely restoration of service continuity \cite{XU2021111642}.

From the resilience curve perspective, the physical dimension defines the severity of the initial impact and how quickly and deeply system performance declines, but does not determine the recovery trajectory, which depends on factors beyond its scope.

\subsection{Operational dimension}
\paragraph{Definition and Importance} The operational dimension refers to the EPS ability to maintain, stabilize, and restore acceptable service levels through coordinated control, decision-making, and resource management following a disturbance. This encompasses real-time and post-event actions such as generation dispatch, network reconfiguration, protection schemes, restoration sequencing, and load prioritization under abnormal conditions \cite{LI2024101592}. Once physical damage has occurred, this dimension determines how available resources are coordinated and how much system flexibility is retained to limit service degradation and support recovery. Remedial actions such as dynamic reconfiguration, islanding, and prioritized restoration are capable of supporting EPS even under restrictive conditions \cite{SHI2021106355}. \looseness=-1

\paragraph{Limitations} Despite this flexibility, adaptive capacity remains constrained by physical asset conditions and the availability of supporting digital infrastructure. Operational resilience assessments are often conducted under fixed network topologies, assuming sufficient flexibility is always available. Such assumptions are unfounded in highly stressed systems or scenarios involving extensive physical damage, where operational actions are limited by asset unavailability and degraded digital support \cite{WANG2022108408, Upadhyay2025Investigate, ONeil2025Enhancing}.

\paragraph{Real-World Implications} These challenges were evidenced during restoration following Hurricane Sandy (2012), which caused extensive damage to transmission and distribution infrastructure. Coordinated measures, including network reconfigurations, critical load prioritization, and strategic crew deployment, enabled partial restoration \cite{DWIVEDI2025125001}, yet their effectiveness was ultimately constrained by the magnitude of physical damage, demonstrating that well-coordinated strategies cannot fully compensate for deficiencies in other resilience dimensions. \looseness=-1

From the resilience curve perspective, the operational dimension primarily shapes the degradation, stabilization, and recovery phases. While it does not prevent initial performance loss, it determines the slope and duration of these trajectories, defining the system's ability to limit impact and restore acceptable operating conditions \cite{WANG2022108408}.

\subsection{Digital-Cyber dimension}

\paragraph{Definition and Importance} The digital-cyber dimension refers to the ability of EPS to sustain observability, coordination, and decision-making through digital and communication infrastructures, encompassing SCADA systems, communication networks, data management architectures, and cybersecurity mechanisms that govern the flow, integrity, and availability of information under both normal and disrupted conditions \cite{Abdelkader2024Securing}. This dimension is a critical enabler of operational resilience, as situational awareness, coordinated control, and timely decisions in highly digitalized EPS rely on the reliability of these infrastructures, which support rapid fault detection, adaptive protection, coordinated restoration, and efficient resource allocation.

\paragraph{Limitations} Despite this enabling role, its contribution remains indirect,i.e, it does not prevent physical damage but conditions how effectively the system responds. When analyzed in isolation, the digital-cyber dimension is often framed as a cybersecurity or information technology (IT) challenge, assuming that stronger defenses or redundancy suffice for resilience. This perspective overlooks interdependencies between digital systems, physical assets, and operational decision-making, underestimating how digital compromises can amplify physical damage, constrain operational responses, and accelerate cascading failures \cite{10238347}.

\paragraph{Real-World Implications} The 2015 cyberattack on the Ukrainian power grid illustrates these interdependencies. Attackers compromised SCADA systems and remotely disconnected multiple substations, causing widespread outages across several regions. The incident demonstrates how a targeted digital breach can amplify physical disruptions, not through equipment damage, but by eliminating operators' ability to monitor, control, and restore the system, transforming a containable event into cascading regional blackouts \cite{XU2021111642}.

From the resilience curve perspective, the digital-cyber dimension primarily influences the degradation, stabilization, and recovery phases. While it does not define the initial physical impact, the integrity of digital infrastructures shapes the slope and continuity of the restoration trajectory, conditioning how efficiently the system stabilizes and restores functionality \cite{XU2021111642}.

\subsection{Climatic-External dimension}
\paragraph{Definition and Importance} The climatic-external dimension refers to the influence of environmental and climatic factors acting as exogenous stressors on EPS' performance and stability, encompassing both acute extreme weather events and long-term climatic trends affecting temperature patterns, precipitation, and resource availability~\cite{GOTEMAN2025115649, Kasimalla2024An, Mehrabanifar2025Enhancing}. This dimension is the primary determinant of system exposure and stress intensity. Climatic conditions define the probability, intensity, and spatial extent of disruptive events, while extreme weather can simultaneously damage physical infrastructure, reduce generation availability, and constrain operational margins. Long-term trends, in turn, progressively erode the system's capacity to maintain adequate levels of flexibility and adequacy.

\paragraph{Limitations} Despite its relevance, this dimension remains inherently exogenous. Namely, its contribution to resilience does not stem from an active system response but from the extent to which EPS can anticipate, absorb, and adapt to externally imposed stressors. When analyzed in isolation, it is typically addressed as a risk factor or statistical characterization of hazards, decoupled from the internal dynamics of the system. This approach focuses on the frequency or severity of events without capturing how climatic stress interacts with endogenous dimensions, leading to underestimation of cumulative effects, prolonged stress conditions, or progressive performance degradation under persistent climatic pressures \cite{MACMILLAN2022112841, Kasimalla2024An}.

\paragraph{Real World Implications} Ecuador's 2024 energy crisis illustrates this limitation. An unprecedented drought severely reduced hydropower generation, the backbone of the national electricity supply, by constraining inflows to the main reservoirs and forcing nationwide rolling blackouts with daily outages of up to 14 hours. This event demonstrates how climatic factors can induce sustained system stress even in the absence of acute physical failures \cite{ecuador_drought_2024}.

From the resilience curve perspective, the climatic-external dimension primarily influences the recovery trajectory and long-term post-event equilibrium. Under persistent climatic stress, system performance may not return to its pre-disturbance level but instead stabilize at a new operational state defined by ongoing environmental constraints.

\begin{table*}[t]
\footnotesize
    \caption{Dimensions of EPS resilience.} \label{tab:core-dimensions-resilience}
    \vspace{4pt}
    
    \newcolumntype{L}{>{\raggedright\arraybackslash}X}
    \newcolumntype{C}{>{\centering\arraybackslash}X}
    \newcolumntype{R}{>{\raggedleft\arraybackslash}X}
    \resizebox{0.99\linewidth}{!}{
    \begin{tabularx}{\textwidth}{p{0.1\textwidth} p{0.18\textwidth} p{0.17\textwidth} X p{0.11\textwidth}}

    \toprule
    \textbf{Dimension} &
    \textbf{Core Function} &
    \textbf{Predominant Phase} &
    \textbf{One-Dimension Limitations} &
    \textbf{Event} \\  
    \midrule
    
    {Physical} &
    Structural robustness and damage absorption &
    Initial impact (robustness / resistance) &
    Overestimates resilience neglecting recovery coordination and system interdependencies &
    Chile earthquake (2010) \\  \hline
    
    Operational &
    Coordinated control, flexibility, and system restoration &
    Degradation, stabilization, and recovery &
    Assumes availability of physical assets and reliable digital support &
    Hurricane Sandy (2012) \\ \hline
    
    Digital--Cyber &
    Digital observability, secure control, and data integrity &
    Degradation, stabilization, and recovery &
    Underestimates cascading physical consequences and operational constraints &
    Ukraine cyberattack (2015) \\ \hline
    
    Climatic--External &
    Hazard exposure and systemic stress intensity &
    Transformative long-term recovery &
    Treated solely as an external hazard, overlooking interaction with system dynamics &
    Ecuador drought crisis (2024) \\ \hline
    
    Economic--Regulatory &
    Institutional incentives and structural investment signals &
    Transformative long-term recovery &
    Disconnects policy design from technical vulnerability and operational realities &
    Texas winter storm (2021) \\ 
    \bottomrule
    \end{tabularx}}
\end{table*}

\subsection{Economic-Regulatory dimension}
\paragraph{Definition and Importance} The economic-regulatory dimension encompasses the institutional, regulatory, and market frameworks within which EPS are planned, operated, and expanded~\cite{YAZDI2024928}. Although it does not directly affect real-time system response, it exerts decisive influence on long-term resilience capacity: regulatory frameworks and market signals determine whether investments in infrastructure reinforcement, resource diversification, and digital modernization are encouraged or discouraged. 

\paragraph{Limitations} This dimension contributes to resilience in an indirect and time-delayed manner. Regulatory and market mechanisms operate over long planning cycles and cannot compensate for immediate physical damage or operational constraints once a disruption occurs. When analyzed in isolation, system response is often reduced to considerations of policy design or market efficiency, disconnected from technical and operational realities, overlooking how regulatory constraints interact with endogenous dimensions \cite{RODRIGUEZMATAS2026115008}. Thus, policies appearing efficient under normal conditions may amplify system fragility during high-impact, low-probability (HILP) events \cite{Schmitz2025Energy}.

\paragraph{Real-World Implications} The Texas electricity market prior to the 2021 winter storm illustrates these dynamics. Operating under an energy-only market design with limited regulatory requirements, the system provided insufficient incentives for generation winterization, reserve procurement, or long-term infrastructure reinforcement, with investment decisions driven by short-term market signals rather than extreme but plausible climatic scenarios. When exceptionally low temperatures struck in February 2021, these institutional constraints resulted in widespread generation outages, severe supply shortages, and large-scale power interruptions, demonstrating how economic and regulatory arrangements can critically constrain resilience even in systems with adequate installed capacity under normal conditions \cite{RODRIGUEZMATAS2026115008}.

From the resilience curve perspective, the economic-regulatory dimension primarily influences the recovery and post-event adaptation phases, determining whether recovery trajectories lead to temporary functional restoration or to deeper structural transformations driven by regulatory reforms, investment strategies, and long-term adaptation policies.

\subsection{Cross-Dimensional Interpretation}
Although each dimension exhibits distinct mechanisms, temporal characteristics, and structural constraints, their roles within the resilience curve reveal common elements, including a core systemic function, a predominant temporal influence, and limitations when assessed in isolation. Table \ref{tab:core-dimensions-resilience} summarizes their core function, dominant resilience phase, principal limitation when assessed in isolation, and representative examples.

\begin{figure}[t]
   \centerline{\includegraphics[trim=0pt 5pt 0pt 0pt, clip, width=1\columnwidth]{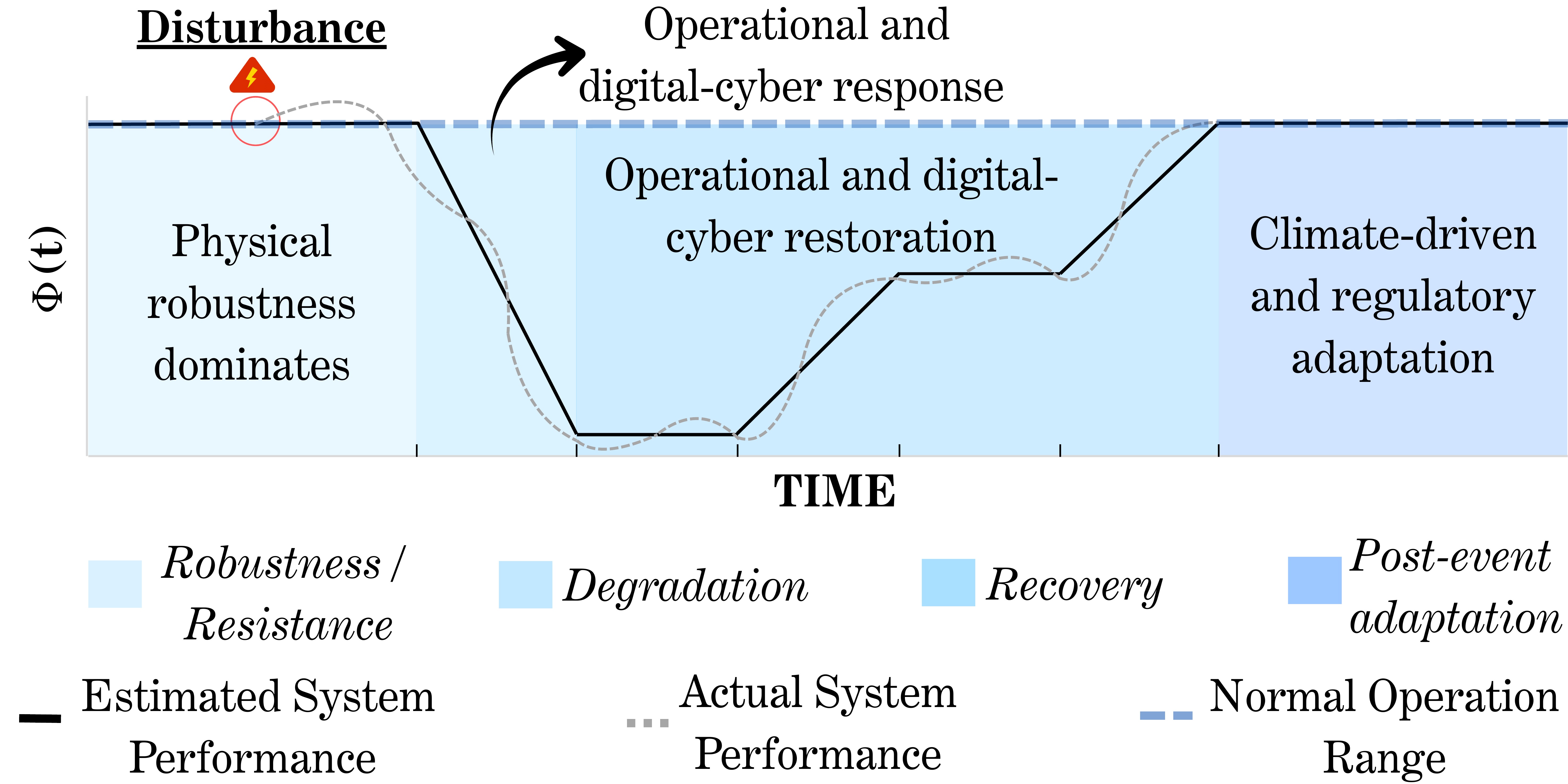}}
    \caption{Resilience curve and dynamic evolution of system performance $\Phi(t)$ across different resilience dimensions.}
    \label{fig:3}
\end{figure}

The dimensions of resilience in EPS do not operate independently nor follow a strictly linear sequence. They interact simultaneously and dynamically, giving rise to resilience trajectories that emerge from the interaction among the different dimensions of the system. Figure~\ref{fig:3} illustrates this behavior through a temporal representation of system performance $\Phi (t)$, in which the phases of initial resistance or robustness, degradation, recovery, and post-event adaptation are articulated.

Immediately after a disturbance, system behavior is conditioned by the physical configuration of the infrastructure. Asset integrity and inherent robustness define the system's immediate resistance capacity and set the baseline from which subsequent dynamics unfold, with the depth and abruptness of performance degradation reflecting the existing structural state of EPS.
As the system transitions from initial impact to degradation, resilience trajectories become increasingly shaped by the interaction between operational flexibility and digital-cyber functionality. Rather than acting sequentially, these dimensions operate in a tightly coupled manner: \emph{i}) operational strategies rely on situational awareness and secure control infrastructures, and \emph{ii}) digital capabilities derive practical meaning from coordinated decision-making. Their combined effectiveness determines whether degradation stabilizes within manageable bounds or escalates into systemic collapse.

The recovery phase extends this interdependence into service restoration, not merely a technical repair process but a coordinated reorganization of system resources, where the pace and continuity of performance restoration depend on how effectively these mechanisms re-establish acceptable operating conditions.
Beyond immediate recovery, long-term adaptation trajectories are shaped by exogenous dimensions that redefine the post-event equilibrium by constraining or enabling structural adjustments. Resilience outcomes are thus determined less by short-term control actions and more by the institutional and environmental context within which the system evolves, implying that resilience does not necessarily mean a return to the pre-event state but the attainment of a new performance equilibrium through continuous interaction among all dimensions.

From this perspective, resilience emerges not from isolated capacities but from complex interactions that evolve over time, a framing that contextualizes the methods and metrics used for its assessment and enables a more coherent interpretation of how existing approaches capture, or fail to capture, the inherent complexity of EPS' resilience.


\section{Resilience Metrics and Limitations Across the Five Dimensions}\label{s:syst}

\subsection{Physical dimension}
\label{ss:physical}
The literature on physical resilience is organized around three dominant paradigms: \emph{i)} distribution planning and resource allocation, \emph{ii)} long-term expansion planning, and \emph{iii)} asset monitoring and lifecycle management.

While these paradigms have advanced infrastructure robustness and long-term adequacy assessment, a structural tension persists across all three. Most planning models rely on steady-state power flows and predefined contingencies, preventing the capture of dynamic damage evolution and cascading failures, while aggregated temporal representations in expansion models limit the modeling of short-term recovery dynamics. Monitoring approaches, though effective at the component level, rarely capture how failures propagate across interconnected networks. More broadly, the widespread assumption of fully available communication infrastructure and coordinated operational control overlooks critical cyber-physical interdependencies during large-scale disruptions, precisely the conditions under which data availability and coordinated control break down. 

As a result, many physical resilience metrics remain proxy indicators that fail to represent dynamic degradation and recovery processes explicitly. Table~\ref{tab:2} systematizes representative studies according to their dimensional scope, types of disturbances, modeling formulations, resilience metrics, mitigation strategies, and methodological limitations.

\subsection{Operational dimension}
\label{ss:operational}
The literature on operational resilience is organized around four dominant paradigms: \emph{i)} post-fault reconfiguration and service restoration, \emph{ii)} security-constrained dispatch and unit commitment, \emph{iii)} service restoration strategies, and \emph{iv)} coordinated distributed energy resources (DER) operation.

While these paradigms have substantially advanced the quantitative characterization of recovery trajectories, load management, and coordination strategies, a structural tension persists across all four. Reconfiguration and scheduling models assume known network states and available coordination, yet disruptive events progressively degrade communication and limit situational awareness. Restoration strategies similarly presume feasible centralized coordination when it is least available, while DER coordination depends on digital synchronization channels that are most vulnerable under major disruptions, rendering mathematically optimal solutions impractical under the conditions they are designed to address. Table~\ref{tab:3} systematizes representative studies according to their dimensional scope, types of disturbances, modeling formulations, resilience metrics, mitigation strategies, and methodological limitations.

\subsection{Digital-Cyber dimension}
\label{ss:cyber}
The literature on digital-cyber resilience is organized around four dominant paradigms: \emph{i)} observability and state estimation, \emph{ii)} communication networks and information and communication technology (ICT) infrastructure, \emph{iii)} cybersecurity and attack mitigation, and \emph{iv)} distributed control.

While these paradigms have advanced the characterization of information flow, situational awareness, and control performance under disrupted conditions, a structural tension persists across all four. Observability and state estimation depend on intact physical networks, while cybersecurity relies on operational communication channels. These assumptions are mutually inconsistent, since cyberattacks degrade the observability needed for their detection, while reduced observability facilitates undetected false data injection. Similarly, ICT models characterize latency and packet loss under nominal conditions, while distributed control inherits those assumptions, rendering its guarantees invalid under correlated disruptions and treating co-degradation across endogenous dimensions as sequential rather than simultaneous. Table~\ref{tab:4} systematizes representative studies according to their dimensional scope, types of disturbances, modeling formulations, resilience metrics, mitigation strategies, and methodological limitations.

\subsection{Climatic-external dimension}
\label{ss:climatic}
The literature on resilience to climatic-external disruptions is organized around three dominant paradigms: \emph{i)} extreme weather disruption, \emph{ii)} climate-driven resource adequacy, and \emph{iii)} compound climate hazards.

While these paradigms have advanced both hazard-constrained operational responses and long-term climate-adaptive planning, a structural tension still remains
across all three. Event-based models analyze acute shocks on nominally intact systems, whereas trend-based models project long-term adequacy under sustained climatic stress; even though extreme events occur on systems already weakened by chronic pressures, mischaracterizing vulnerability when degraded systems respond differently to superimposed shocks. Compound hazard approaches partially bridge this gap by modeling co-occurring stressors, but still assume unchanged operational response capacity and intact coordination infrastructure.

Table~\ref{tab:5} systematizes representative studies according to their dimensional scope, types of disturbances, modeling formulations, resilience metrics, mitigation strategies, and methodological limitations.

\subsection{Economic-regulatory dimension}
\label{ss:economic}
The literature on economic-regulatory resilience is organized around three dominant paradigms: \emph{i)} investment incentives and planning; \emph{ii)} electricity market design; and \emph{iii)} regulatory governance and public policy frameworks.

While these paradigms have advanced understanding of how market signals, capacity mechanisms, and institutional frameworks shape long-term resilience capacity, a structural tension persists among the three. Investment planning derives resilience capacity from market signals generated by designs that already assume available infrastructure, creating a circular dependency in which resilience is required yet systematically undervalued, so that economically rational decisions may yield systems structurally unprepared for extreme events, particularly when HILP conditions are underrepresented. A further tension arises between governance and investment: regulatory frameworks assume stable implementation conditions while overlooking the physical and operational constraints that determine feasibility, producing a persistent disconnect between economic optimization and system behavior under disruption.

Table~\ref{tab:table6}  systematizes representative studies according to their dimensional scope, types of disturbances, modeling formulations, resilience metrics, mitigation strategies, and methodological limitations.

\subsection{Cross-Dimensional Synthesis and Structural Gaps}
\label{ss:synthesis}
 
\begin{figure}[t]
   \centerline{\includegraphics[width=1\columnwidth]{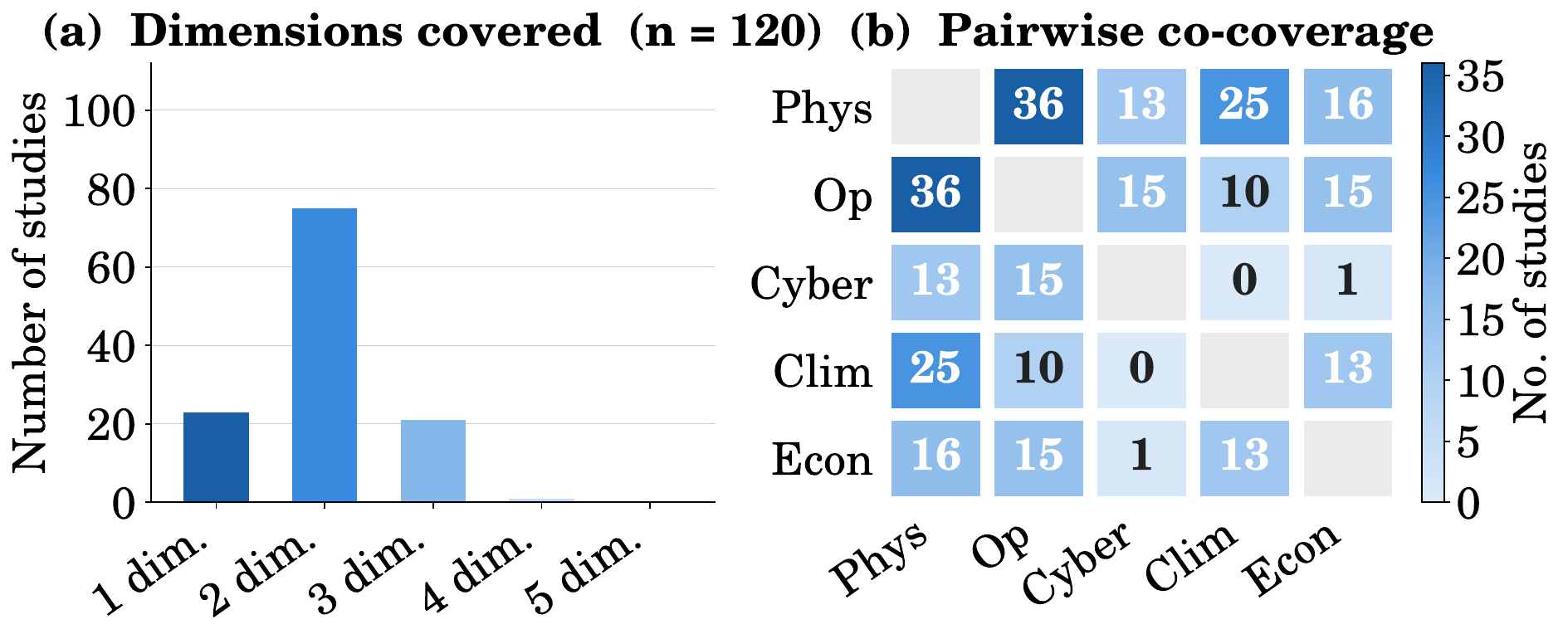}}
    \caption{(a)~Dimensional coverage distribution and (b)~pairwise co-coverage of the reviewed literature.}
    \label{fig:4}
\end{figure}

Figure~\ref{fig:4} maps the reviewed studies against the five resilience dimensions of the proposed framework, providing a quantitative basis for the observations that emerge from Sections~\ref{ss:physical}--\ref{ss:economic}. 

\paragraph{Dimensional fragmentation and structural asymmetry}
The literature reveals a strong fragmentation in resilience assessment across dimensions. Of the 120 studies reviewed, 62\% consider only two resilience dimensions, 19\% focus on a single dimension, and 18\% address three; only one study reaches four dimensions, while none integrates all five simultaneously. Moreover, multidimensional coverage is highly asymmetric and concentrated around a limited set of interactions, primarily the physical--operational pair, followed by operational--cyber couplings. Interactions beyond these dominant combinations remain largely unexplored, with no studies addressing the cyber--climatic relationship and only one examining the cyber--economic pair.

These patterns expose three systematic gaps:

\begin{itemize}

\item Physical constraints bound operational decisions, and cyber constraints shape coordination; yet the bidirectional feedback loop, where physical damage degrades cyber situational awareness, driving suboptimal decisions that further amplify physical damage, remains unmodeled as a closed-loop process.

\item Event-based approaches model acute shocks on nominally intact systems, while trend-based approaches project long-term adequacy under aggregated stress scenarios; their interaction, how sustained climatic pressure reshapes the system that acute shocks then strike, remains unmodeled.

\item Economic mechanisms, whether in investment planning or market design, are evaluated assuming intact physical and operational infrastructure; how those mechanisms behave when that assumption fails — under physical damage, climatic stress, or cyber disruption — remains systematically unexamined.

\end{itemize}

Across all five dimensions, the reviewed literature exhibits three recurring structural assumptions that constrain its analytical scope: coordination infrastructure is assumed continuously available, failures are modeled as isolated rather than correlated and progressive disturbances, and resilience dimensions are treated independently. As a result, most existing formulations focus on individual dimensions or adjacent pairs, failing to capture the simultaneous compromise and mutual amplification that compound stress can induce across interconnected dimensions.

\section{Multidimensional Resilience Assessment Under Cyber-Physical Attack Scenarios}
\label{s:Comp}

The cross-dimensional analysis in Section~\ref{ss:synthesis} reveals three systematic gaps: bidirectional co-degradation, acute–chronic climatic interaction, and limited economic integration under disruption. To address these gaps, this section first develops the $\mathcal{MDRI}$; a quantitative index that aggregates cross-dimensional degradation under simultaneous endogenous and exogenous stress interactions and then applies it to a case study of escalating cyber-physical attack scenarios \cite{romero2026resilience}.

\subsection{System Performance and Resilience Loss}

The system performance function $\Phi(t)$ combines frequency deviation and inter-machine coherency, as defined in Eq.~\eqref{eq:1}:

\begin{equation}
\Phi(t) = w_f\Phi_f(t) + w_s\Phi_s(t)
\label{eq:1}
\end{equation}

\begin{equation*}
\begin{aligned}
\Phi_f(t) &= \max\!\left(0,\; 1 - \frac{\left|f_{\mathrm{COI}}(t) - f_{\mathrm{nom}}\right|}
{\left|f_{\mathrm{nom}} - f_{\mathrm{crit}}\right|}\right), 
\\[6pt]
\Phi_s(t) &= \max\!\left(0,\; 1 - \frac{\Delta f_{\mathrm{gen}}(t)}
{\Delta f_{\mathrm{gen}}^{\mathrm{coh}}}\right)
\end{aligned}
\end{equation*}

\noindent where $f_{\mathrm{COI}}(t)$ is the center-of-inertia frequency, 
$f_{\mathrm{nom}}$ and $f_{\mathrm{crit}}$ are the nominal and critical frequencies, 
$\Delta f_{\mathrm{gen}}(t)=\max_{i,j}|f_i(t)-f_j(t)|$ is the inter-generator frequency spread, and $\Delta f_{\mathrm{gen}}^{\mathrm{coh}}$ is the coherency tolerance band. The weights satisfy $w_f,w_s\in[0,1]$ and $w_f+w_s=1$. Equal weights are adopted since frequency deviation and coherence loss are considered equally important stability phenomena.

The resilience loss metric $R_{\mathrm{loss}}$ quantifies cumulative performance degradation over a horizon $T_H$ relative to the pre-disturbance operating point. Integration starts at the disturbance time $t_{0,i}$ for scenario $S_i$. To account for collapse, the truncated performance function $\tilde{\Phi}_i(t)$ is defined as

\begin{equation*}
\tilde{\Phi}_i(t) = 
\begin{cases} 
\Phi_i(t), & t \leq t_{\mathrm{lim}} \\ 
0, & t > t_{\mathrm{lim}}
\end{cases}
\end{equation*}

\noindent where $t_{\mathrm{lim}}=t_f$ if the system recovers within $T_H$, and $t_{\mathrm{lim}}=t_{\mathrm{col}}$ if collapse occurs at $t_{\mathrm{col}}$. The resilience loss is then computed as

\begin{equation}
R_{\mathrm{loss},i} = \frac{1}{\Phi_{0,i}} \int_{t_{0,i}}^{t_{0,i}+T_H} 
\max\!\left(0,\; \Phi_{0,i} - \tilde{\Phi}_i(t)\right) dt
\label{eq:2}
\end{equation}

\noindent where $\Phi_{0,i}$ denotes the pre-disturbance performance level for scenario $S_i$.

\subsubsection{Physical Dimension}

The physical disruption index $D_{\mathrm{phy},i}$ in Eq.~\eqref{eq:3} quantifies the weighted loss of generation capacity caused by a disturbance in scenario $S_i$:

\begin{equation}
D_{\mathrm{phy},i} = \sum_{r \in R} \omega_r \frac{P_{\mathrm{r,lost},i}}
{P_{\mathrm{r,total}}}
\label{eq:3}
\end{equation}

\noindent where $P_{\mathrm{r,lost},i}$ is the disconnected or unavailable capacity of resource $r$ in scenario $S_i$ (MW), $P_{\mathrm{r,total}}$ is its installed capacity (MW), $\omega_r\in[0,1]$ is the assigned weight, and $R=\{\mathrm{PV,\, sync,\, storage,\, substations,\dots}\}$ denotes the set of resource types.

The weights reflect the relative importance of affected resources in $S_i$, while unaffected types are excluded. When multiple resources are impacted, weights are assigned according to their criticality (i.e., inertia, reserves, or critical-load support). If only one resource drives the disruption, $\omega_r=1$.

\subsubsection{Operational Dimension}
The operational disruption index $D_{\mathrm{op},i}$ characterizes the system dynamic response to a disturbance in scenario $S_i$ by combining frequency variation rate, performance degradation, and generator coherency, as defined in Eq.~\eqref{eq:4}:

\begin{equation}
D_{\text{op},i} = \frac{X_i}{1 + X_i}, \quad 
X_i = \frac{1}{3}\left( \frac{|\text{RoCoF}|_{\max,i}}{\text{RoCoF}_{\text{crit}}} 
+ \frac{\delta_{\Phi,i}}{\delta_{\Phi}^{\text{crit}}} 
+ \frac{\Delta f_{\text{gen},i}^{\max}}{\Delta f_{\text{gen}}^{\text{crit}}}\right)
\label{eq:4}
\end{equation}

\begin{equation*}
\delta_{\Phi,i} =
\frac{\Phi_{0,i} - \Phi_{\text{nadir},i}}{\Phi_{0,i}}
\end{equation*}

\noindent where $|\mathrm{RoCoF}|_{\max,i}$ is the maximum absolute rate of change of frequency, $\delta_{\Phi,i}$ is the normalized performance drop, $\Phi_{\mathrm{nadir},i}$ is the minimum value of $\Phi(t)$ during the event, and $\Delta f_{\mathrm{gen},i}^{\max}=\max_t \Delta f_{\mathrm{gen},i}(t)$ is the peak inter-generator frequency spread. Equal weights are assigned to the three terms due to their complementary role in transient stability degradation. Critical thresholds are set to $\mathrm{RoCoF}^{\mathrm{crit}}=1.0$~Hz/s \cite{entsoe2023inertia}, $\delta_{\Phi}^{\mathrm{crit}}=0.05$, and $\Delta f_{\mathrm{gen}}^{\mathrm{crit}}=2.0$~Hz.

The saturating form bounds $D_{\text{op},i}$ within $[0,1)$ while preserving sensitivity near the critical region ($X_i=1 \to D_{\text{op},i}=0.5$). As $X_i$ increases, $D_{\text{op},i}$ asymptotically approaches unity.

\subsubsection{Digital-Cyber Dimension}
The digital-cyber disruption index $D_{\mathrm{cyb},i}$ quantifies the impact of a disturbance on system observability and controllability during scenario $S_i$, incorporating observability, controllability, integrity, and availability into the normalized index of Eq.~\eqref{eq:5}:

\begin{equation}
D_{\mathrm{cyb},i} = \sum_{j \in \mathcal{K}} w_{\mathrm{cy}_j} 
\cdot \frac{N_{\mathrm{compr}_j,i}}{N_{\mathrm{scope}_j}}
\label{eq:5}
\end{equation}

\noindent where $\mathcal{K}=\{\mathrm{obs,\, ctrl,\, int,\, av}\}$ denotes the evaluated cyber aspects, $N_{\mathrm{compr}_j,i}$ is the number of compromised assets in aspect $j$ for scenario $S_i$, $N_{\mathrm{scope}_j}$ is the total number of evaluated assets, and $w_{\mathrm{cy}_j}\geq0$ are weighting factors satisfying $\sum_{j \in \mathcal{K}} w_{\mathrm{cy}_j}=1$.

\subsubsection{Climatic-External Dimension}

The climatic-external disruption index $D_{\mathrm{clim},i}$ represents environmental stressors that exacerbate system degradation during scenario $S_i$. Multiple climatic factors are aggregated into the normalized index of Eq.~\eqref{eq:6}:

\begin{equation}
D_{\mathrm{clim},i} = \sum_{k \in \mathcal{C}} w_{c_k} \cdot I_{c_k,i}
\label{eq:6}
\end{equation}

\noindent where $C=$\{temperature, snow, wind, ice, humidity, extreme weather, \dots\} is the set of climatic stressors considered, $I_{c_k,i}\in[0,1]$ is the normalized intensity of stressor $k$ in scenario $S_i$, and $w_{c_k}\geq0$ are weighting factors satisfying $\sum_{k \in \mathcal{C}} w_{c_k}=1$.

\subsubsection{Economic-Regulatory Dimension}
\label{ss:EcReg}
The economic-regulatory disruption index $D_{\mathrm{econ\text{-}reg},i}$ comprises two sub-components that are combined through a weighted aggregation. The economic sub-index $D_{\mathrm{econ},i}^{\mathit{und}}$ quantifies how underinvested the system is in preventive controls relative to the realized cost of failure, defined as Eq.~\eqref{eq:7}:

\begin{equation}
D_{\mathrm{econ},i}^{\mathrm{\textit{und}}} = 1 - \exp\!\left( 
-\frac{C_i^{\mathrm{to\text{-}cost}}}{C_i^{\mathrm{pr\text{-}inv,min}}} 
\right)
\label{eq:7}
\end{equation}

\noindent where $C_i^{\mathrm{pr\text{-}inv,min}}$ is the minimum preventive investment that would have prevented or mitigated scenario $S_i$, and $C_i^{\mathrm{to\text{-}cost}}$ is the total realized 
cost, decomposed as:

\begin{equation*}
C_i^{\mathrm{to\text{-}cost}} = \mathrm{VoLL} \cdot \mathrm{ENS}_i + 
C_i^{\mathrm{rep}} + C_i^{\mathrm{pen}} + C_i^{\mathrm{isl}}
\end{equation*}

\noindent where $\mathrm{VoLL}$ is the Value of Lost Load (monetary unit per MWh), $\mathrm{ENS}_i$ is the energy not supplied during $S_i$ (MWh), $C_i^{\mathrm{rep}}$ is the repair and replacement cost of damaged equipment, $C_i^{\mathrm{pen}}$ is the cost of regulatory penalties incurred for reliability standard violations, and $C_i^{\mathrm{isl}}$ is the islanding cost incurred during $S_i$, which captures the operational cost of either proactive or reactive island formation, both of which are captured additively in $C_i^{\mathrm{isl}} = C_i^{\mathrm{isl,proac}} + C_i^{\mathrm{isl,reac}}$.

Complementarily, the regulatory sub-index $D_{\mathrm{reg},i}^{\mathit{vul}}$ quantifies institutional vulnerabilities by measuring the fraction of reference control strategies that are absent or non-compliant in the context where the scenario $S_i$ occurs, as given by Eq.~\eqref{eq:8}:

\begin{equation}
D_{\mathrm{reg},i}^{\mathit{vul}} = \frac{1}{N_v^{\mathrm{ref}}} 
\sum_{j=1}^{N_v^{\mathrm{ref}}} v_{j,i}
\label{eq:8}
\end{equation}

\noindent where $v_{j,i} \in \{0,1\}$ indicates the presence ($v_{j,i} = 1$) or absence ($v_{j,i} = 0$) of a regulatory weakness in control category $j$, and $N_v^{\mathrm{ref}}$ is the total number of controls in the reference framework.

The comprehensive $D_{\mathrm{econ\text{-}reg},i}$ index is expressed in Eq.~\eqref{eq:9}.

\begin{equation}
D_{\mathrm{econ\text{-}reg},i} = 
 w_e \cdot D_{\mathrm{econ},i}^{\mathit{und}}
+w_r \cdot D_{\mathrm{reg},i}^{\mathit{vul}} 
 + w_c \cdot \left(  
D_{\mathrm{econ},i}^{\mathit{und}} 
D_{\mathrm{reg},i}^{\mathit{vul}}\right)
\label{eq:9}
\end{equation}

\noindent where $w_e$, $w_r$, and $w_c$ denote the weights of the economic, regulatory, and coupling terms, respectively, with $w_e + w_r + w_c = 1$. We impose three principles: \emph{(i)} equal standalone importance, $w_e = w_r$, reflecting no prior preference between financial underinvestment and institutional fragility; \emph{(ii)} dominance of standalone terms, $w_e + w_r > w_c$, so the coupling term modulates rather than dominates $D_{\mathrm{econ\text{-}reg},i}$; and \emph{(iii)} a strictly positive coupling weight, $w_c > 0$, to capture their documented synergy.

\subsection{Multidimensional Resilience Index}

The proposed formulation rests on the premise that evaluating resilience dimensions \textit{independently} underestimates system-wide impacts, since simultaneous compromise of multiple aspects and their interdependencies produces degradations undetectable by single-dimensional assessments \cite{Sun2024Scenario, 10221865}. The $\mathcal{MDRI}_i$ is therefore constructed in two layers: an endogenous core $\mathcal{M}(S_i;\gamma_i)$ capturing coupled degradation among physical, operational, and digital-cyber dimensions; and exogenous amplifiers scaling that degradation proportionally to climatic and economic-regulatory conditions.

\noindent\textbf{Endogenous Core:} Let $\mathcal{K}_{\mathrm{sim}} = \{\mathrm{phy,\, op,\, cyb}\}$ 
be the set of endogenous dimensions. The endogenous core decomposes degradation into an \textbf{additive} and a \textbf{coupling} contribution, as given by eq.~\eqref{eq:10}:

\begin{equation}
\mathcal{M}(S_i;\gamma_i) =
\underbrace{%
  \frac{1}{|\mathcal{K}_{\mathrm{sim}}|}
  \sum_{k \in \mathcal{K}_{\mathrm{sim}}} D_{k,i}
}_{\overline{D}_i \; (\text{additive})}
+\;
\gamma_i
\underbrace{%
  \prod_{k \in \mathcal{K}_{\mathrm{sim}}} D_{k,i}
}_{\Pi_i \; (\text{coupling})}
\label{eq:10}
\end{equation}

\noindent where $D_{k,i} \in [0,1]$ is the normalized sub-index of dimension $k$ in scenario $S_i$, with equal weights $1/|\mathcal{K}_{\mathrm{sim}}|$. The additive term 
$\bar{D}_i$ measures mean severity across dimensions independently. The coupling term $\Pi_i = \prod_{k} D_{k,i}$ captures additional degradation arising from simultaneous 
cross-dimensional compromise; it collapses to zero whenever any single dimension remains uncompromised ($D_{k,i} \to 0$), and reaches its maximum only when all dimensions are 
jointly and severely degraded, encoding the cascading failure mechanism whereby an intact dimension suppresses impact propagation, while simultaneous degradation across 
all dimensions produces mutual amplification beyond the additive prediction.

\noindent\textbf{Regime Assignment and Coupling Parameter:} The parameter $\gamma_i$ takes one of two values. Scenarios driven by a single disturbance vector are assigned 
to the \textbf{additive regime} ($\gamma_i = 0$), under which $\mathcal{M}(S_i;0) = \bar{D}_i$ and no cross-dimensional interaction is considered. Scenarios where all 
endogenous dimensions are simultaneously compromised are assigned to the \textbf{coupled regime} ($\gamma_i = 1$), activating $\Pi_i$ and amplifying degradation beyond the additive baseline. Unity is the structurally neutral value at which $\Pi_i$ and $\bar{D}_i$ contribute on the same scale, and its selection is further supported by a ranking-invariance property: for any scenario set in which both $\bar{D}_i$ and $\Pi_i$ are monotonically ordered, the strict inequality $\bar{D}_i + \gamma_i\Pi_i < \bar{D}_j + \gamma_i\Pi_j$ holds for all $\gamma_i \geq 0$, 
so that $\gamma_i$ governs the \textit{magnitude} of coupling amplification but not the comparative ordering of scenarios, a property verified numerically in Section~\ref{ss:Robust}.
 
\noindent\textbf{MDRI Formulation:} The $\mathcal{MDRI}$ for scenario $S_i$ is given by Eq.~\eqref{eq:11}:
 
\begin{equation}
\mathcal{MDRI}_i = \mathcal{M}(S_i;\gamma_i) \cdot
\prod_{j \in \mathcal{K}_{\mathrm{ext}}} (1 + D_{j,i})
\label{eq:11}
\end{equation}
 
\noindent where $\mathcal{K}_{\mathrm{ext}} = \{\mathrm{clim,\, econ\text{-}reg}\}$ is the set of exogenous dimensions and $D_{j,i} \geq 0$ is the normalized sub-index of exogenous dimension $j$ in scenario $S_i$. Each factor $(1 + D_{j,i})$ amplifies the endogenous core proportionally to the exogenous stress level. When no exogenous stress is present ($D_{j,i} = 0$), the factor reduces to unity, and the index simplifies to $\mathcal{MDRI}_i = \mathcal{M}(S_i;\gamma_i)$. Since the exogenous amplifiers act multiplicatively on the endogenous core, $\mathcal{MDRI}_i$ is not normalized to $[0,1]$ and serves as a comparative metric across scenarios rather than an absolute resilience score.
 
The multiplicative structure $\prod_{j \in \mathcal{K}_{\mathrm{ext}}}(1+D_{j,i})$ encodes two structural assumptions. First, exogenous dimensions act as mutually independent stress amplifiers rather than autonomous sources of degradation; each factor scales the endogenous core independently, such that climatic and economic-regulatory conditions may jointly amplify endogenous degradation without directly altering one another. Second, the form $(1+D_{j,i})$ ensures that the absence of exogenous stress $(D_{j,i}=0)$ recovers the endogenous-only formulation, i.e., $\mathcal{MDRI}_i=\mathcal{M}(S_i;\gamma_i)$.

\subsection{Case Study and Attack Scenarios} 
\label{s:system_models}

\begin{figure}[t]
   \centerline{\includegraphics[trim=0pt 35pt 0pt 35pt, clip, width=1\columnwidth]{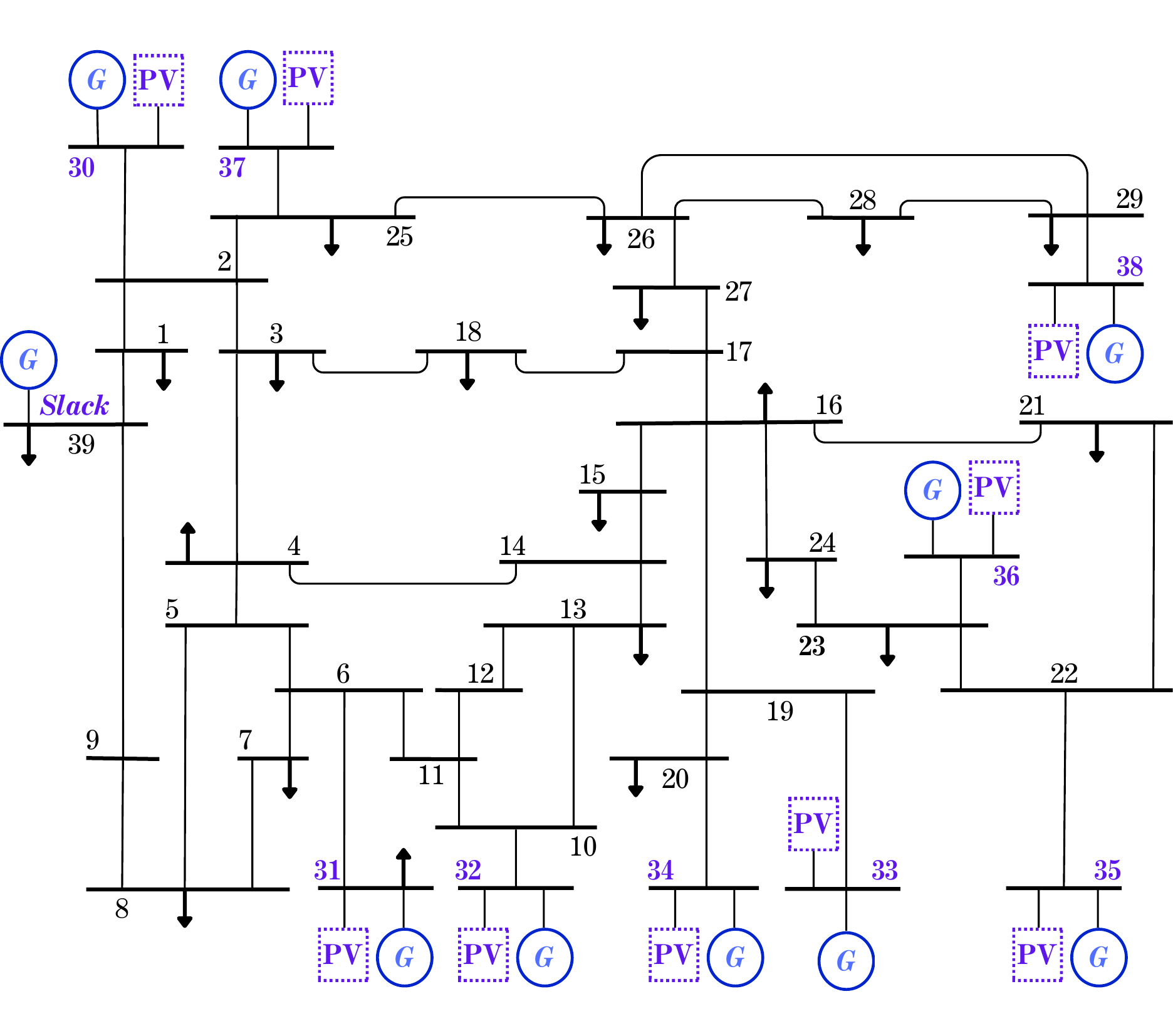}}
    \caption{IEEE 39-bus system with PV plants deployed across multiple buses.}
    \label{fig:39bus}
\end{figure}

\noindent\textbf{System Model:~} The proposed framework is validated on the IEEE 39-bus New England test system implemented in MATLAB/Simulink. The original system includes 10 synchronous generators with a total installed capacity of 10,610~MW. To represent increasing inverter-based resource penetration, 1,500~MW of synchronous generation is replaced by nine utility-scale grid-forming PV plants connected at buses $30-38$, corresponding to 14.1\% of total capacity as shown in Figure~\ref{fig:39bus}.

\noindent\textbf{Attack Scenarios:~} Three escalating attack scenarios are evaluated based on the coordinated cyberattack on the Polish energy infrastructure on December 29, 2025~\cite{CERTPolska2026}. Cyber-physical attacks are considered because they simultaneously stress the physical, operational, and digital-cyber dimensions.

\noindent\textbf{Threat Model:~} A state-sponsored adversary targeting both IT and OT domains is assumed, consistent with the tactics, techniques, and procedures (TTPs) attributed to the ELECTRUM/Sandworm group~\cite{CERTPolska2026, Dragos2026, ESET2026}. The attacker is assumed to possess prior knowledge of grid topology and ICS protocols, exploiting exposed perimeter devices and default credentials to gain initial access and move laterally into OT networks. The attack includes coordinated multi-vector actions across distributed sites, including control signal manipulation and communication disruption.

\textbf{Scenario 1} -- \textbf{Single-plant baseline attack}:~A control input attack manipulates the active power reference of the PV plant at Bus 33 (190~MW), selected for its proximity to the highest-load buses in the system. As a result, a generation loss at this bus produces measurable system-wide frequency transients while remaining within the single-plant scope for our baseline scenario. Operators retain full communication and control over all the remaining plants. This represents an isolated cyberattack, without coupled interactions among endogenous dimensions or additional climatic and regulatory considerations, serving as a baseline for comparisons with the following scenarios.

\textbf{Scenario 2} -- \textbf{Multi-vector cascading attack}: A coordinated attack replicates TTPs documented in the Polish grid incident: \emph{i)} communication disruption targeting six PV plants, eliminating operator observability and control, and \emph{ii)} forced disconnection of 1,115~MW of PV capacity (74.3\% of total). The targeted plants (Buses 31, 32, 34, 35, 37, and 38) are geographically dispersed.
To reflect the elevated winter demand observed during the Polish incident, a $25\%$ load increase is introduced as an operational stress factor. Sub-zero temperatures and snowstorms are not modeled separately, as their effects are already captured by this increase. Scenario~2 thus isolates the impact of multi-vector endogenous compromise without exogenous amplification.

\textbf{Scenario 3} -- \textbf{Realistic state-sponsored attack}: This use case represents the worst-case scenario and assumes a state-sponsored attacker with unlimited resources. All PV plants are targeted simultaneously, resulting in a complete loss of PV generation capacity (1,500~MW). Climatic conditions reflect a concurrent extreme weather event~\cite{CERTPolska2026, Dragos2026}, economic stressors are explicitly incorporated, and the regulatory environment enforces maximum penalties.

\subsection{Attack Impact and Dynamic Response}
\begin{figure}[t]
   \centerline{\includegraphics[, width=1\columnwidth]{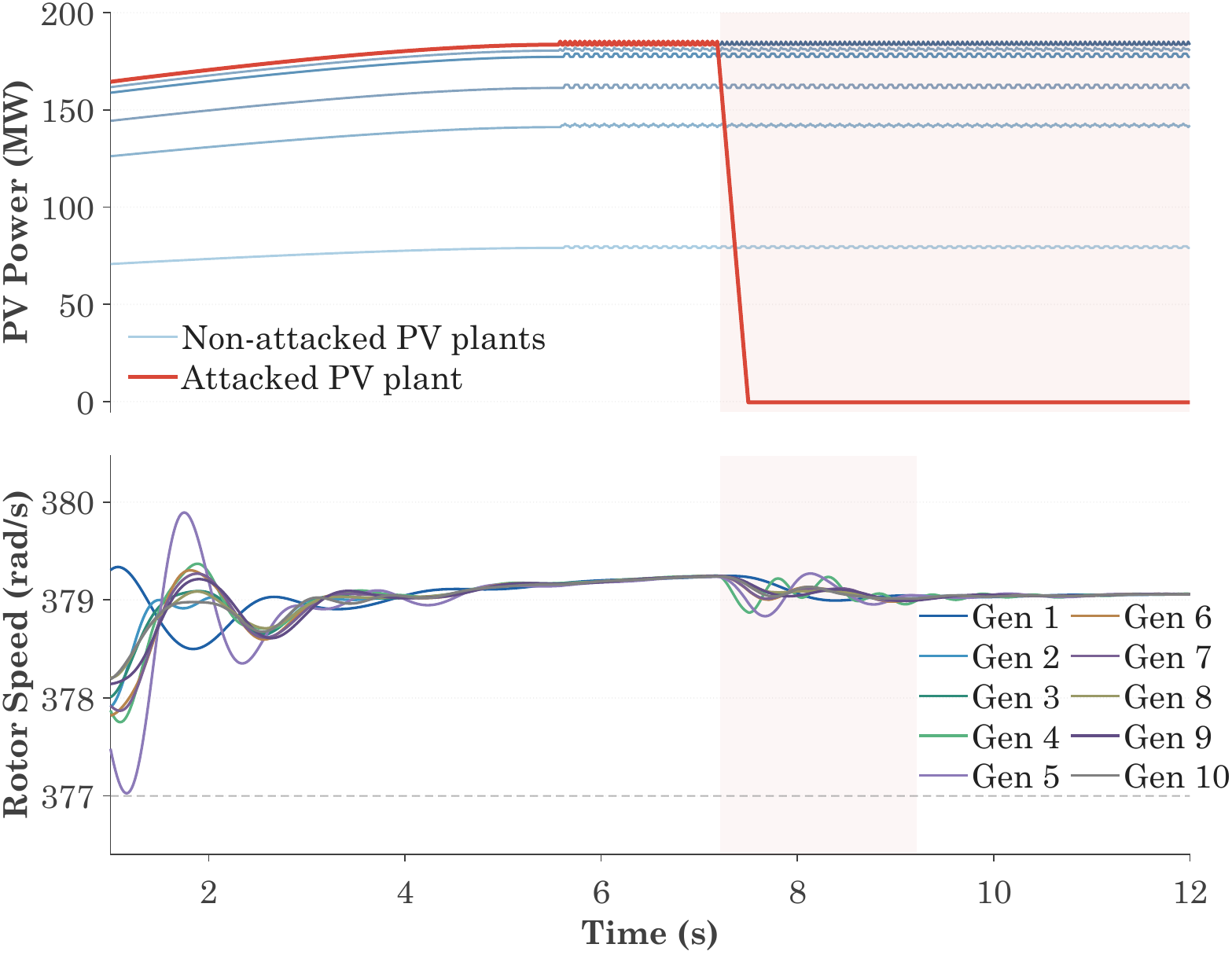}}
    \caption{PV generation and rotor speed response for Scenario 1.}
    \label{fig:scenario1}
\end{figure}

\begin{figure}[t]
   \centerline{\includegraphics[, width=1\columnwidth]{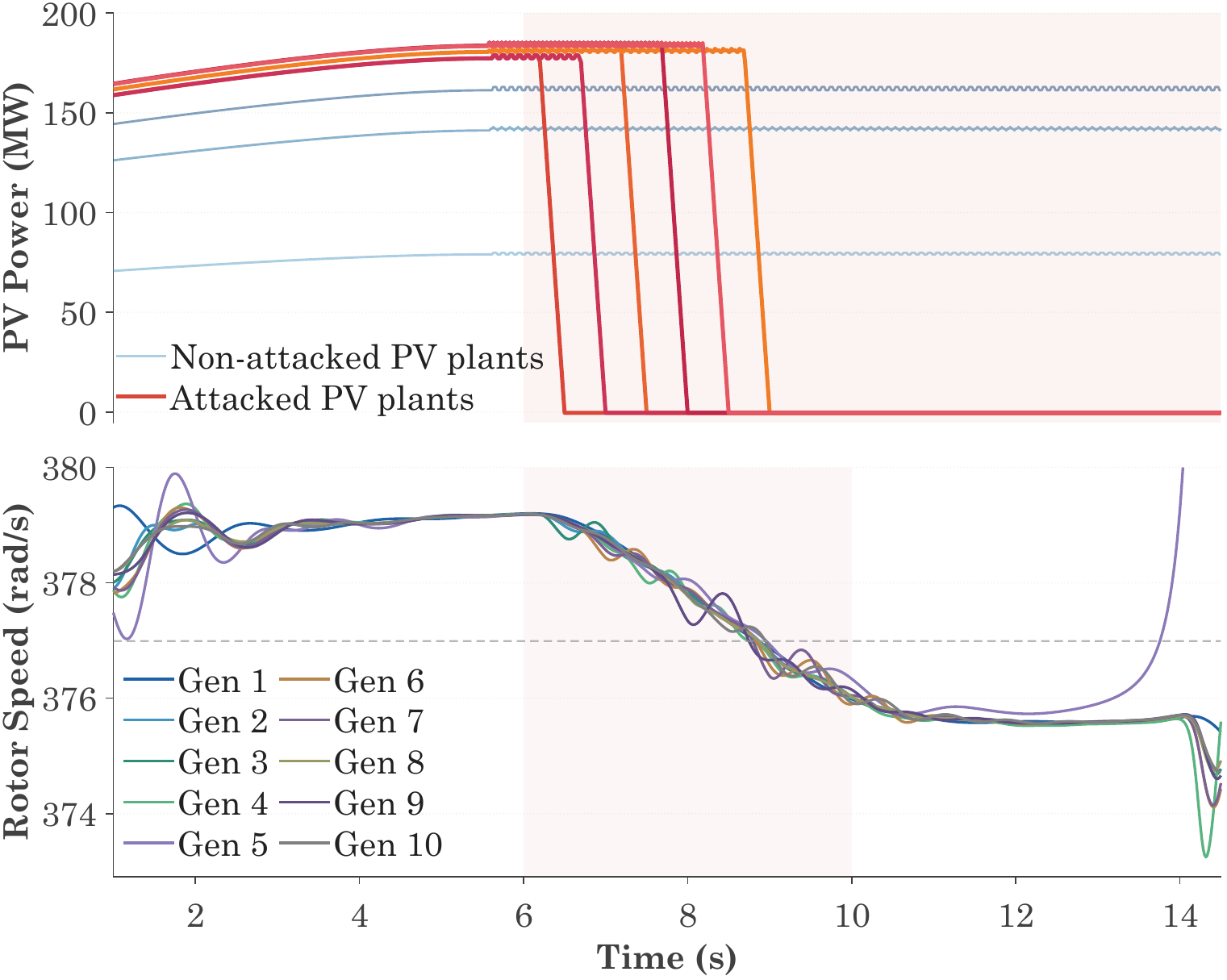}}
    \caption{PV generation and rotor speed response for Scenario 2.}
    \label{fig:scenario2}
\end{figure}

\begin{figure}[t]
   \centerline{\includegraphics[, width=1\columnwidth]{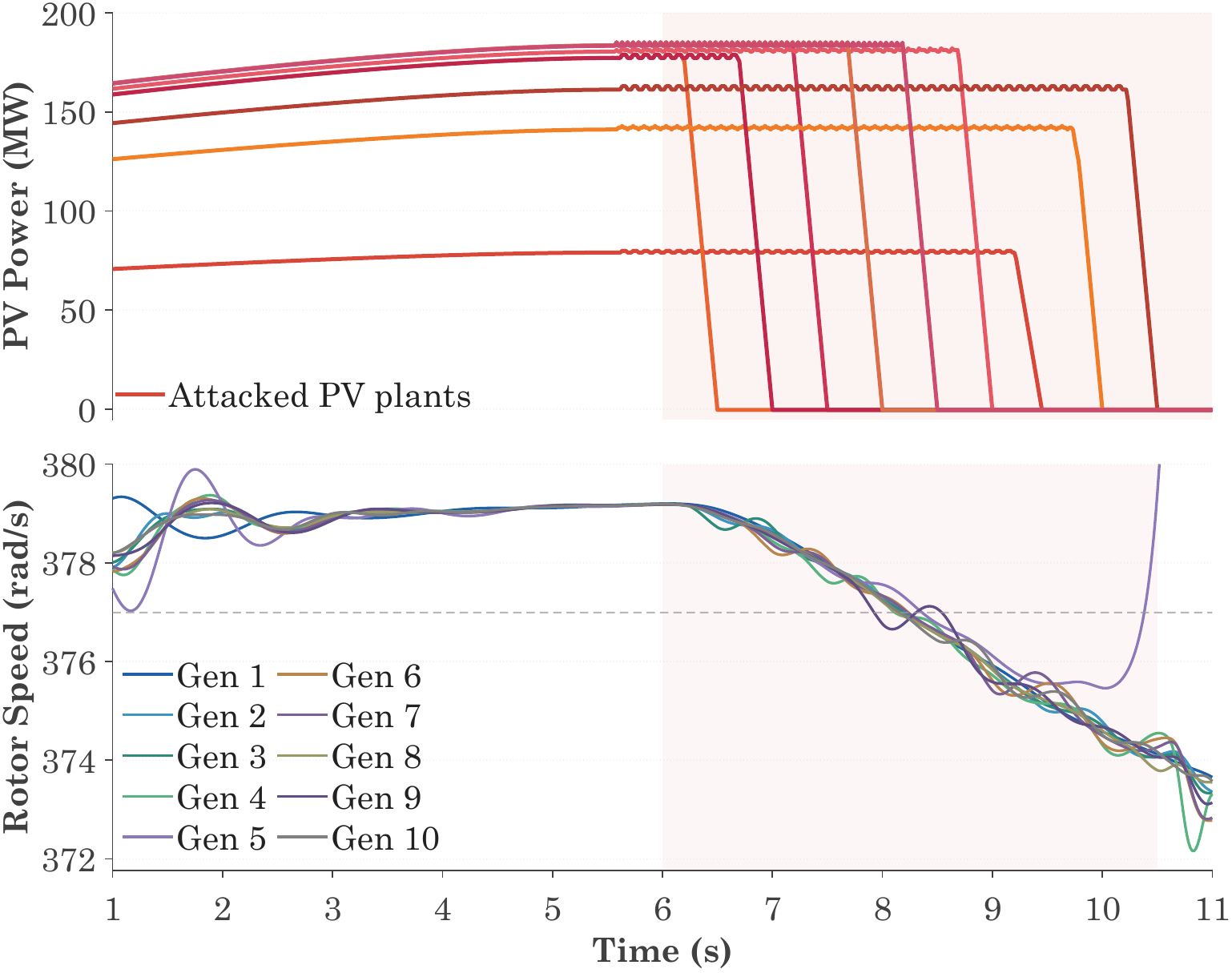}}
    \caption{PV generation and rotor speed response for Scenario 3.}
    \label{fig:scenario3}
\end{figure}

Figures~\ref{fig:scenario1}--\ref{fig:scenario3} present the PV generation output and synchronous generator rotor speeds for each scenario, while Table~\ref{tab:freq} summarizes the corresponding frequency response.

\begin{table}[t]
\footnotesize
\centering
\caption{Frequency response metrics comparison.} 
\begin{tabular}{l c c c}
\toprule
{Metric} & {Scenario 1} & {Scenario 2} & {Scenario 3} \\
\midrule
Nadir frequency [Hz]        & 60.323    & 59.776  & 59.544  \\
Time to frequency nadir [s] & 8.720      & 12.820   & 10.950  \\
Maximum RoCoF [Hz/s]        & -0.055    & -0.089  & -0.282  \\
RoCoF time [s]              & 7.500       & 6.500     & 10.817  \\
Steady-state frequency [Hz] & 60.332    & \textit{unstable} & \textit{unstable} \\
$\Delta f_{\mathrm{gen}}$ [Hz] & 0.00012 & 3.097   & 3.115   \\
\bottomrule
\end{tabular}
\vspace{0.3em}
\parbox{0.9\linewidth}{\footnotesize
\textit{Note:} The larger RoCoF time in Scenario 3 is caused by angular instability. Maximum RoCoF occurs due to rotor speed divergence.
}
\label{tab:freq} 
\end{table}

\noindent \textbf{Scenario 1:}~As demonstrated in Figure~\ref{fig:scenario1}, the attack on the PV plant forces its output from 190~MW to zero (at $t=7s$), while non-attacked plants maintain nominal generation. All generators exhibit brief transient oscillations before converging to a steady state. During Scenario 1, the system maintains frequency stability with negligible frequency deviations.

\noindent\textbf{Scenario 2:}~As demonstrated in Figure~\ref{fig:scenario2}, six PV plants are simultaneously disconnected, removing 1,115~MW from the system. The rotor speed responses reveal dynamic instabilities, generators attempt a coordinated response, but beyond $t\approx8s$ their trajectories diverge, leading to loss of synchronism.
During Scenario 2, frequency performance deteriorates, with a 61.8\% increase in maximum RoCoF compared to Scenario 1. The generator frequency fluctuations reach $3.097$Hz, indicating a complete loss of synchronism among synchronous machines. 

\noindent\textbf{Scenario 3:}~As demonstrated in Figure~\ref{fig:scenario3}, nine PV plants are simultaneously disconnected, removing 1,500~MW from the system. The rotor speed responses reveal a rapid and unrecoverable dynamic collapse. Generators initially attempt a coordinated response, but beyond $t\approx7s$ their trajectories diverge, leading to loss of synchronism.
During Scenario 3, frequency performance deteriorates most severely among all evaluated cases. The maximum RoCoF reaches $-0.282$~Hz/s, representing a 413\% increase with respect to Scenario 1 and a 217\% increase with respect to Scenario 2. Similarly, the generator frequency spread reaches $\Delta f_{\mathrm{gen}} = 3.115$~Hz, exceeding the $3.097$~Hz observed in Scenario 2. 

\begin{figure}[t]
   \centerline{\includegraphics[, width=1\columnwidth]{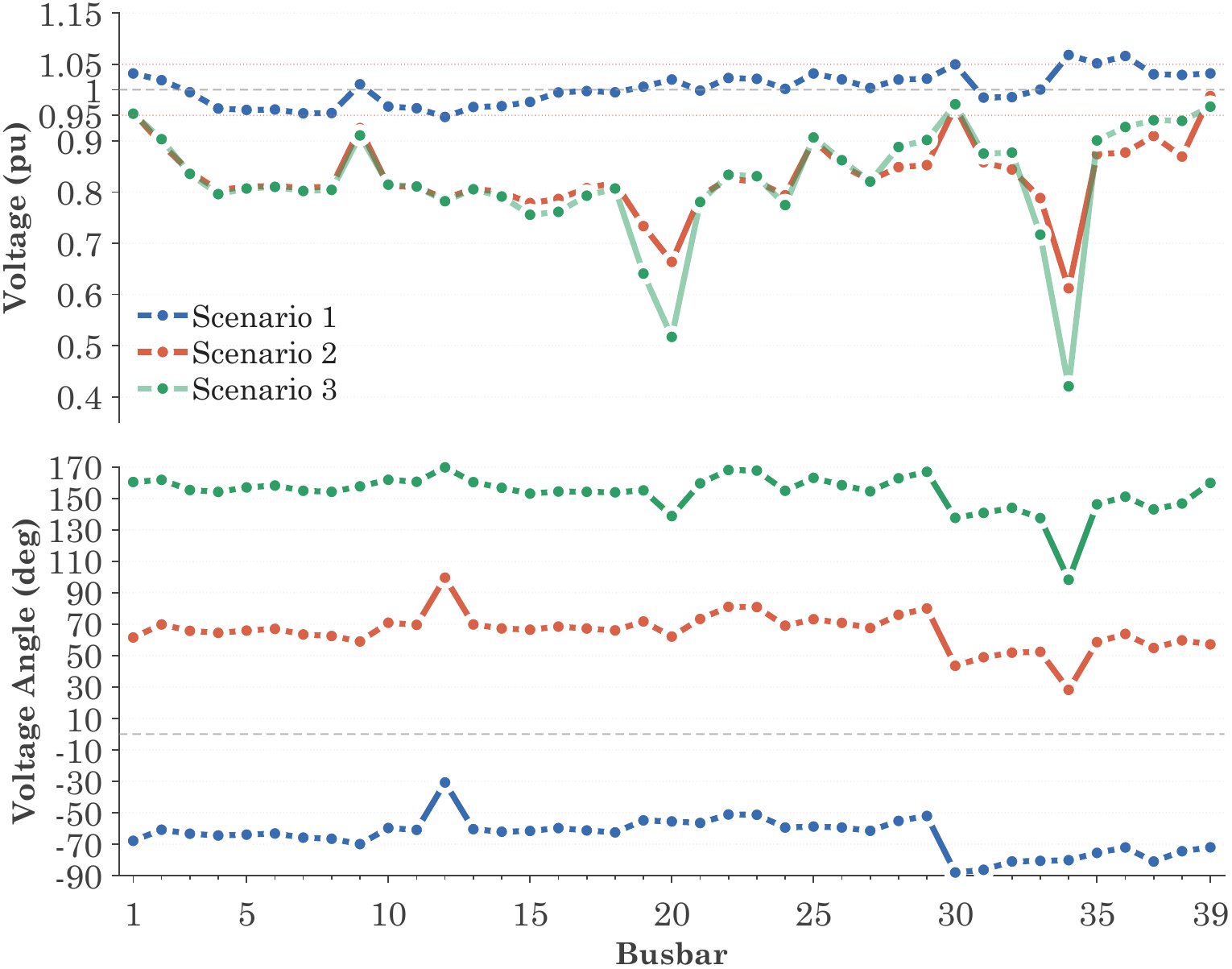}}
    \caption{Voltage magnitudes and angle profiles across all 39 buses at $t=15s$.}
    \label{fig:volt_and_angles}
\end{figure}

While bus voltages remain within nominal ranges and no load shedding is triggered in scenario 1, scenarios 2 and 3 exhibit widespread voltage sags and substantial increases in angle dispersion, reflecting the loss of angular coherency across the system  (Figure~\ref{fig:volt_and_angles}). \looseness=-1

\subsection{System Performance and Resilience Curves}

\begin{figure}[t]
   \centerline{\includegraphics[, width=1\columnwidth]{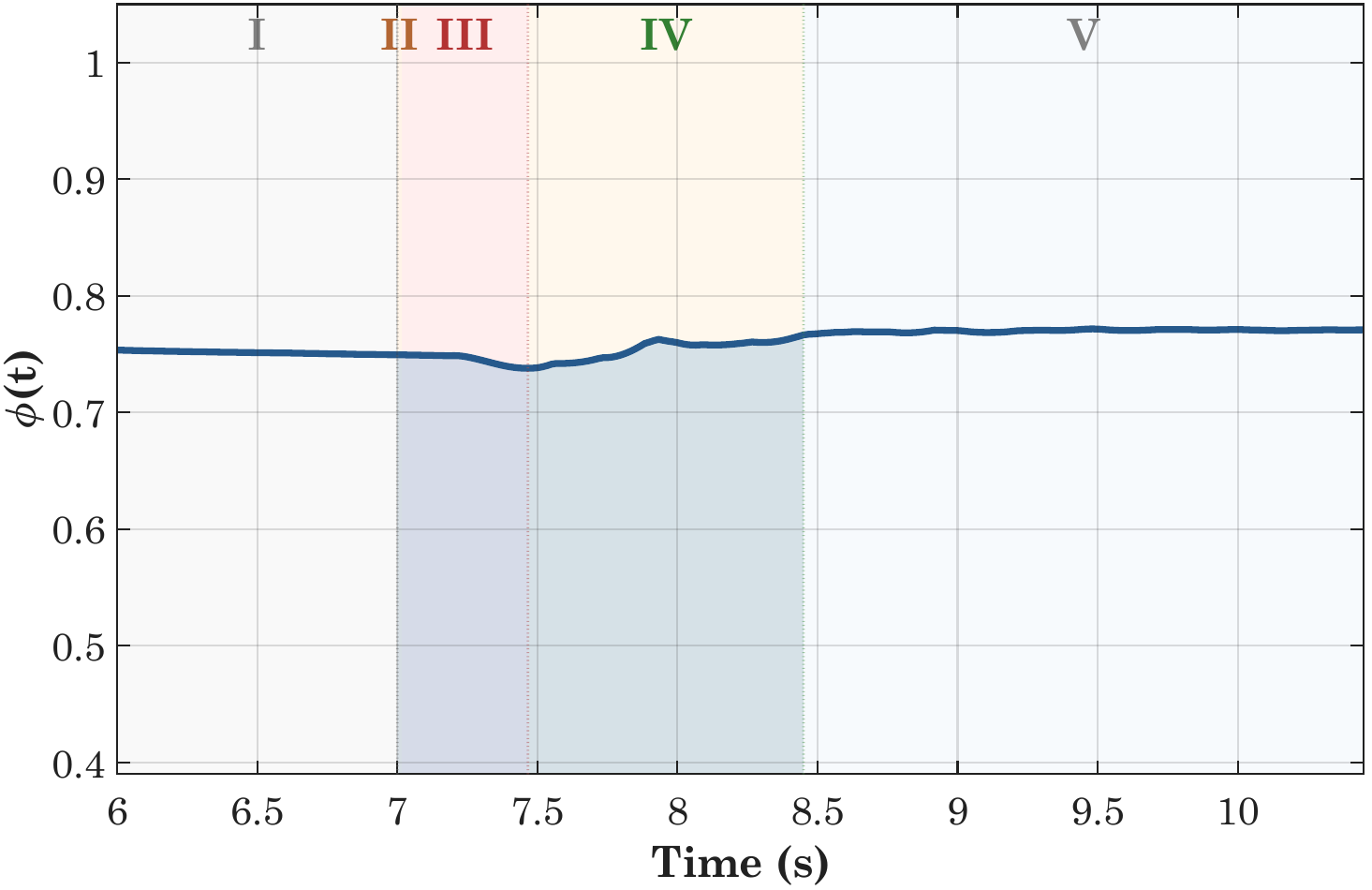}}
    \caption{System performance function and phases for Scenario 1. }
    \label{fig:sysperf1}
\end{figure}

\begin{figure}[t]
   \centerline{\includegraphics[, width=1\columnwidth]{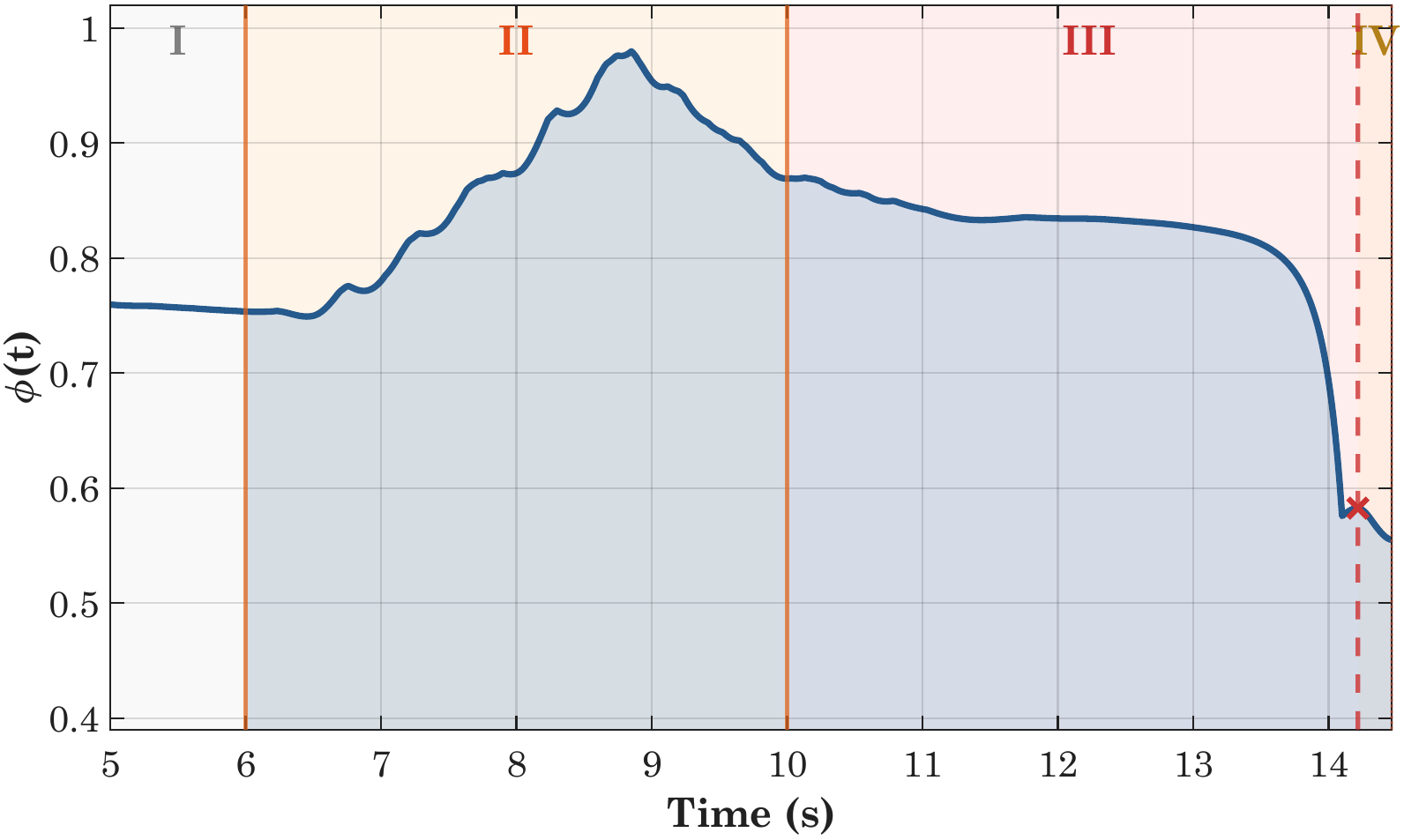}}
    \caption{System performance function and phases for Scenario 2. }
    \label{fig:sysperf2}
\end{figure}

\begin{figure}[t]
   \centerline{\includegraphics[, width=1\columnwidth]{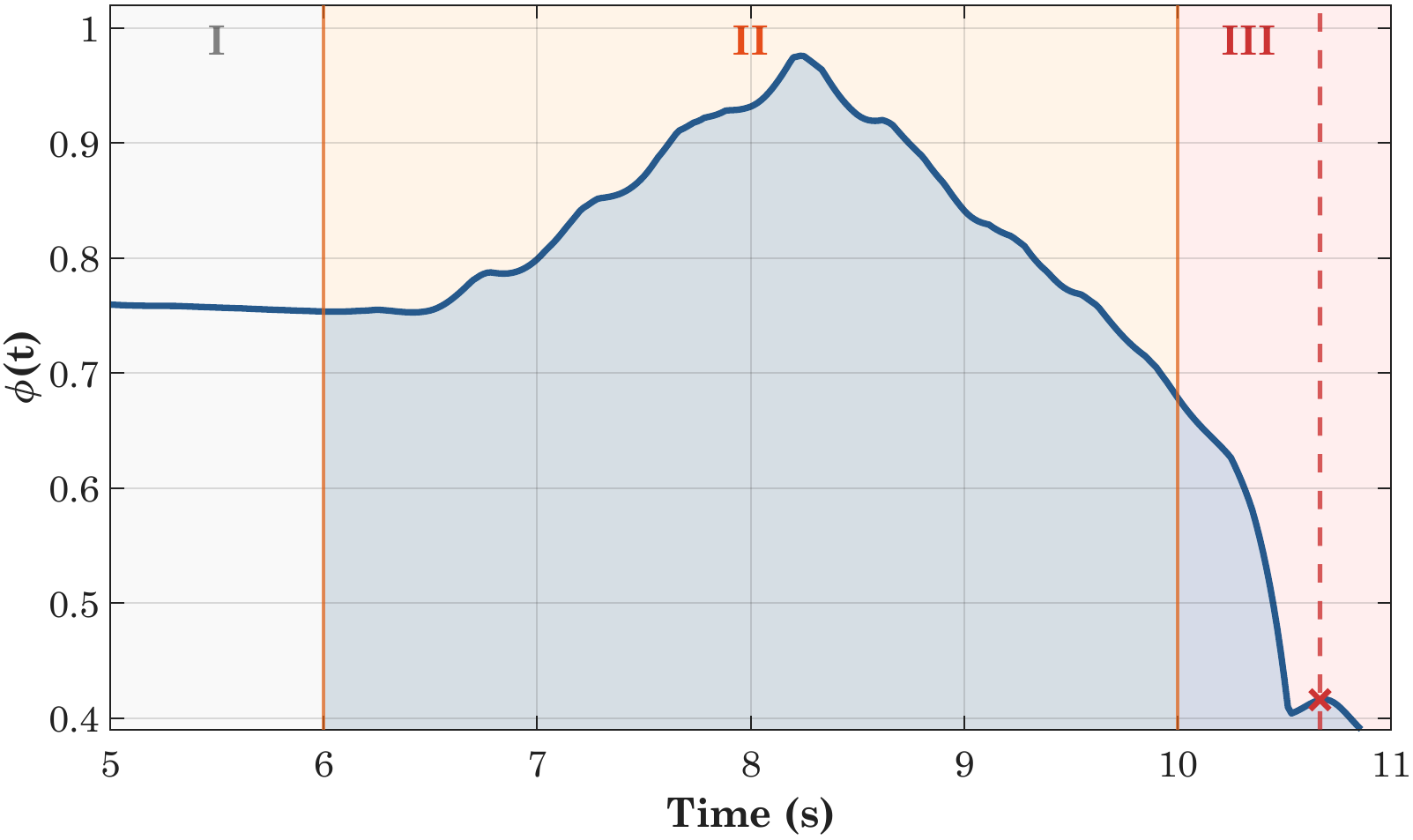}}
    \caption{System performance function and phases for Scenario 3.}
    \label{fig:sysperf3}
\end{figure}

Figures~\ref{fig:sysperf1}--\ref{fig:sysperf3} present the system performance \(\Phi(t)\) for scenarios 1, 2, and 3 (\(w_f\!=\!w_s\!=\!0.5\), 
\(f_{\mathrm{nom}}\!=\!60\)Hz, \(f_{\mathrm{crit}}\!=\!59\)Hz, \(\Delta f_{\mathrm{gen}}^{\mathrm{coh}}\!=\!1.0\)Hz, \(T_H\!=\!15\)s), where $\Phi_0(t)\!<\!1$ due to residual PV-integration deviations.

\textbf{Scenario 1} -- Successful Recovery:
Following the attack on a single PV plant, the system performance degrades from $\Phi_0 = 0.7674$ to a nadir of $0.7379$ at $t = 7.47s$, representing a 3.8\% drop. Recovery to $\Phi = 0.7670$ (99.9\% of pre-event level) occurs within $1.45s$, and $R_{\text{loss}} =0.0305$. The system performance function exhibits five distinct phases: \emph{i)} pre-event steady state, \emph{ii)} absorption ($0.02s$), \emph{iii)} degraded operation ($0.45s$), \emph{iv)} recovery ($0.98s$), and \emph{v)} post-event equilibrium.

\textbf{Scenario 2} -- Cascading System Collapse:
The multi-vector attack triggers progressive degradation from $\Phi_0 = 0.7674$ to a nadir of $0.5549$ at $t = 12.82s$, a 27.7\% loss. The system does not recover and the collapse is detected at $t = 14.22s$ with $\Phi = 0.5824$, and $R_{\text{loss}} =1.0080$. The system performance exhibits only four phases (absent post-event equilibrium): \emph{i)} pre-event steady state, \emph{ii)} attack propagation ($4.00s$), during which the system attempts to absorb the generation loss over approximately $3s$ before losing stability entirely, \emph{iii)} degraded operation ($4.22s$), \emph{iv)} failed recovery ($0.35s$). The short-lived phase \emph{iv} fails to mitigate the cascading instability, leading to complete system collapse at $t=14.22s$.

\begin{figure}[t]
   \centerline{\includegraphics[, width=1.07\columnwidth]{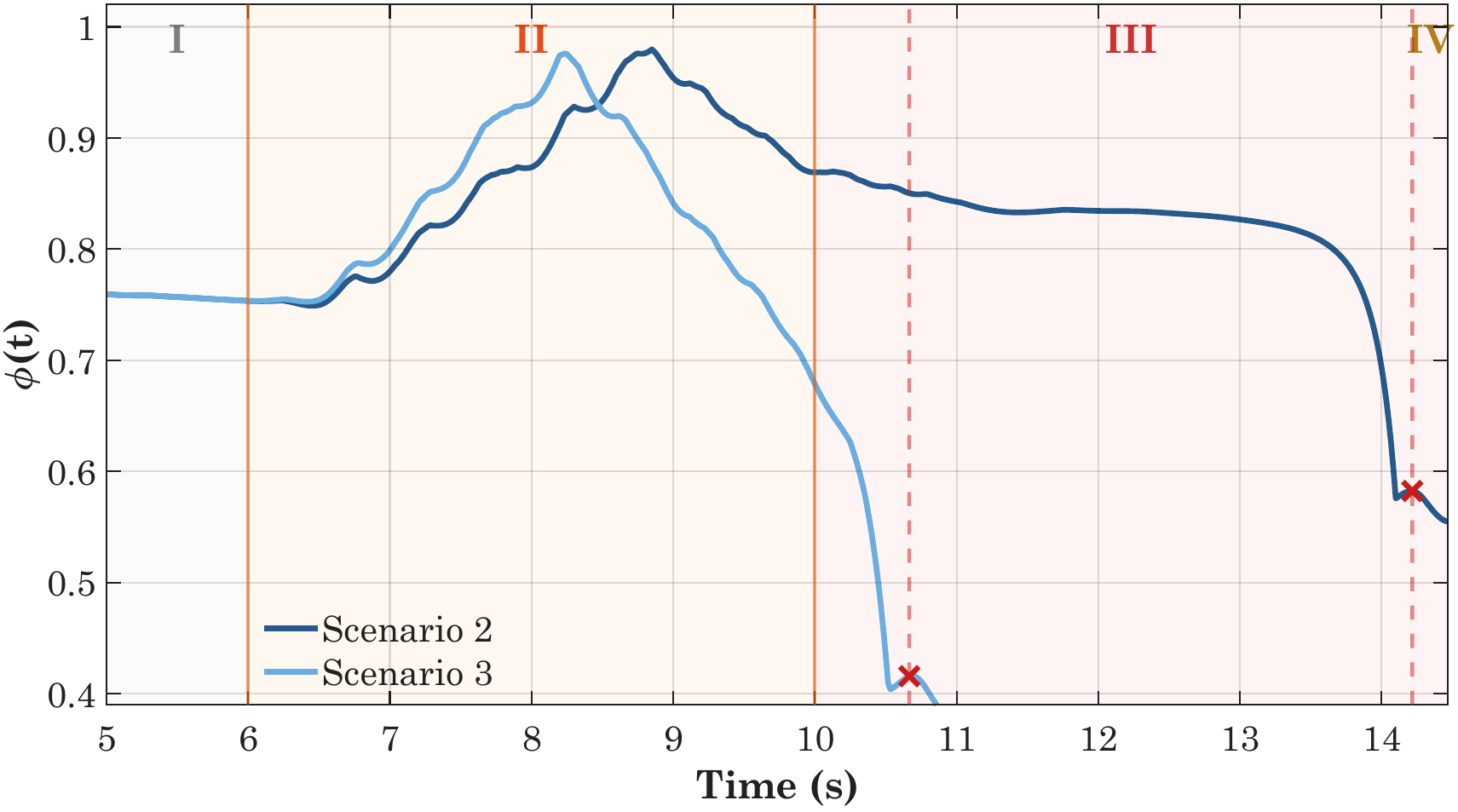}}
    \caption{System performance function between scenarios 2 and 3.}
    \label{fig:syscomp}
\end{figure}

\textbf{Scenario 3} -- Accelerated Catastrophic Collapse:
The simultaneous disconnection of nine PV plants triggers an immediate and unrecoverable degradation from $\Phi_0 = 0.7674$ to a nadir of $0.3809$ at $t = 10.67s$, a 50.3\% loss, the most severe among all evaluated scenarios. 
The system collapses at $t = 10.95s$ with $\Phi = 0.4158$, and $R_{\text{loss}} = 4.7193$. The system performance exhibits only three phases (absent both recovery and post-event equilibrium): \emph{i)} pre-event steady 
state, \emph{ii)} attack propagation ($4.5s$), during which the system is overwhelmed with no meaningful absorption capacity, and \emph{iii)} degraded operation ($0.95s$), collapsing within $0.05s$ of reaching the nadir, leaving no time for any corrective response, confirming that tripling the number of attacked plants reduces the time to instability. Figure~\ref{fig:syscomp} illustrates the system performance comparison between scenarios 2 and 3. 

\subsection{Multidimensional Indices Calculation}

\subsubsection{Physical Dimension}
The physical sub-index $D_{\text{phy},i}$ measures the fraction of inverter-based generation capacity lost under scenario $S_i$. Since all scenarios exclusively affect the PV fleet, Eq.~\eqref{eq:3} reduces to
$D_{\text{phy},i} = \frac{P_{\text{PV,lost},i}}{P_{\text{PV,total}}}$ with $\omega_{\text{PV}} = 1$.
This yields $D_{\text{phy},S1} = 0.127$, $D_{\text{phy},S2} = 0.743$, and $D_{\text{phy},S3} = 1.000$.

\subsubsection{Operational Dimension}
The operational sub-index $D_{\text{op},i}$ is computed using Eq.~\eqref{eq:4}. In Scenario~1, all indicators remain below their critical thresholds, yielding $X_{S1}=0.272$ and $D_{\text{op},S1}=0.214$. In Scenarios~2 and~3, the simultaneous disconnection of multiple plants causes both $\delta_{\Phi}$ and $\Delta f_{\text{gen}}^{\max}$ to exceed their limits. Specifically, $\delta_{\Phi,S2}=0.277$ and $\delta_{\Phi,S3}=0.504$ exceed $\delta_{\Phi}^{\text{crit}}$ by factors of 5.5 and 10.1, respectively, while the inter-generator frequency spread reaches $3.097$~Hz and $3.115$~Hz, surpassing the 2.0~Hz coherency threshold. Thus, $X_{S2}=2.393$ and $X_{S3}=3.973$, both within the super-critical regime ($X>1$). Eq.~\eqref{eq:4} maps these values to $D_{\text{op},S2}=0.705$ and $D_{\text{op},S3}=0.799$, preserving discrimination between cascading and catastrophic degradation.

\subsubsection{Digital-Cyber Dimension}
The digital-cyber sub-index $D_{\text{cyb},i}$ is computed via Eq.~\eqref{eq:5} across four aspects: observability, controllability, integrity, and availability, each weighted equally at $w_{\text{cy}_j} = 0.25$, reflecting their complementary and non-redundant roles in characterizing operator situational awareness and system controllability.
The scope is set to $N_{\text{scope}_j} = 9$ for all aspects, corresponding to the nine PV plants comprising the inverter-based fleet. The number of compromised assets per aspect and scenario is determined from the attack~\cite{CERTPolska2026,Dragos2026} and are summarized in Table~\ref{tab:assets_compromised}.

\begin{table}[t]
\footnotesize
\centering
\caption{Cyber asset compromise mapping per aspect and scenario.}
\begin{tabularx}{\linewidth}{lXccc}
\toprule
{Aspect} & {Justification} & {S1} & {S2} & {S3} \\
\midrule
Observability  
    & Remote monitoring loss
    & $0/9$ & $6/9$ & $9/9$ \\
Controllability 
    & Remote control loss     
    & $0/9$ & $6/9$ & $9/9$ \\
Integrity       
    & Signal/firmware compromise            
    & $1/9$ & $6/9$ & $9/9$ \\
Availability    
    & Disconnection of attacked plants      
    & $1/9$ & $6/9$ & $9/9$ \\
\bottomrule
\end{tabularx}
\label{tab:assets_compromised}
\end{table}

In Scenario~1, the attack manipulates the active power reference of a single plant without disrupting operator communications, leaving observability and controllability fully intact across the fleet.
Only integrity and availability are partially compromised at the targeted bus, yielding $D_{\text{cyb},S1} = 0.056$. In Scenario~2, the coordinated disruption of communications at six plants eliminates remote observability and controllability at those sites, while firmware corruption and forced disconnection extend integrity and availability compromise to the same six units, giving $D_{\text{cyb},S2} = 0.667$. Scenario~3 assumes a state-sponsored adversary with unlimited resources targeting all nine plants simultaneously, saturating all four aspects and yielding $D_{\text{cyb},S3} = 1.000$.

\subsubsection{Climatic-External Dimension}

The climatic-external sub-index $D_{\text{clim},i}$ captures exogenous environmental stress. No climatic effects are considered for Scenarios~1 and~2 ($D_{\text{clim},1} = D_{\text{clim},2} = 0$). Scenario~3 incorporates an extreme winter event affecting Polish power infrastructure, considering four stressors: temperature, snow load, wind, and ice. All variables are normalized with respect to standard design reference values illustrated in Table \ref{tab:climatic}. Equal weights $w_k = 0.25$ are assigned to all four stressors, reflecting that any single climatic mechanism dominates infrastructure failure under extreme winter conditions in the Polish context,  thus, $D_{\text{clim},S3} = 0.765$. 

\begin{table}[t]
\scriptsize
\centering
\caption{Climatic stressor normalization for Scenario~3.}
\begin{tabular}{lcccc}
\toprule
Stressor & Normalization & Event value & Reference value & Intensity \\
\midrule
Temperature & $\dfrac{|T_{\text{event}}|}{|T_{\text{ref}}|}$
  & $-25$~\textdegree{}C~\cite{Worlddata2024}
  & $-30$~\textdegree{}C~\cite{Eurelectric2024}
  & $0.833$ \\[8pt]
Snow load & $\min\!\left(\dfrac{s_{\text{event}}}{s_{k}},1\right)$
  & $2.4$~kN/m$^{2}$~\cite{EN1991_AnnexD}
  & $2.0$~kN/m$^{2}$~\cite{EN1991_Annex_C}
  & $1.000$ \\[8pt]
Wind & $\dfrac{v_{\text{event}}}{v_{\text{ref}}}$
  & $25$~m/s~\cite{Ribeiro2017}
  & $28$~m/s~\cite{Chmielewski2023}
  & $0.893$ \\[8pt]
Ice & $\dfrac{t_{\text{event}}}{t_{\text{ref}}}$
  & $10$~mm~\cite{Tomaszewski2019}
  & $30$~mm~\cite{CIGRE2006}
  & $0.333$ \\
\bottomrule
\end{tabular}
\label{tab:climatic} 
\end{table}

\subsubsection{Economic-Regulatory Dimension}

As with the climatic-external dimension, the economic-regulatory sub-index $D_{\mathrm{econ\text{-}reg},i}$ is applied exclusively to Scenario~3, which represents the worst-case state-sponsored threat necessitating maximum regulatory enforcement, and accordingly, $D_{\text{econ-reg},S1} =D_{\text{econ-reg},S2} = 0$.

\paragraph{Economic sub-index} The economic sub-index $D_{\mathrm{econ},i}^{\mathrm{\textit{und}}}$ is computed through Eq.~\eqref{eq:7}, which requires the total realised cost $C_{i}^{\text{to-cost}}$ and the minimum preventive investment $C_{i}^{\text{pr-inv,min}}$. The Value of Lost Load for Poland is not officially established~\cite{ACER2022}. However, following the range reported across European Union member states of \euro{}1,500-22,940/MWh~\cite{Swinand2019,ACER2020}, a conservative value of $\text{VoLL} = \text{\euro{}}8{,}000$/MWh is adopted. 

\textit{Energy not supplied} ($C_{\text{ENS},S3}$): In scenario~3, the system collapses at $t_{\text{col}} = 10.67$~s with an approximate operating load of 6{,}000~MW, yielding $\text{ENSs}_3 = 17.78 \text{ MWh}$, thus, $C_{\text{ENS},S3} \approx \text{\euro{}}0.14 \text{ M}.$

\textit{Equipment repair and replacement} ($C_{S3}^{\text{rep}}$): Nine PV plants are compromised including their inverter, RTU, and SCADA assets. Based on unit replacement benchmarks for utility-scale PV inverters and OT infrastructure in the 50--200~MW range~\cite{IRENA2024}, an estimate of \euro{}3.5~M per plant yields $C_{S3}^{\text{rep}} = \text{\euro{}}31.5 \text{ M}.$

\textit{Regulatory penalties} ($C_{S3}^{\text{pen}}$): Under the European cybersecurity framework, Polish energy operators qualify as essential entities under the NIS2 Directive ~\cite{NIS2Directive2022}. Non-compliance with mandatory cybersecurity risk-management obligations may result in administrative fines of up to \euro{}10~M or 2\% of total worldwide annual turnover, whichever is higher. Adopting the fixed ceiling conservatively, a penalty of $C_{S3}^{\text{pen}} = \text{\euro{}}10$~M is assumed.

\textit{Islanding cost} ($C_{S3}^{\text{isl}}$): Both proactive island formation and reactive black-start restoration are considered. Based on ENTSO-E restoration cost benchmarks for systems of comparable installed capacity~\cite{ENTSOE2023}, a total cost of \euro{}5.5~M is assumed, comprising \euro{}2.0~M for proactive islanding and \euro{}3.5~M for sequential black-start restoration.

The minimum preventive investment $C_{S3}^{\text{pr-inv,min}}$ covers full cybersecurity hardening of nine plants, encompassing multi-factor authentication deployment, firmware lifecycle management, IT/OT network segmentation, OT intrusion detection systems, personnel security training, and supply-chain risk auditing. Drawing on ENISA cost benchmarks for OT security deployment in the energy sector~\cite{ENISA2023}, an estimate of \euro{}1.2~M per plant yields $C_{S3}^{\text{pr-inv,min}} = \text{\euro{}}10.8$~M.

The total realized cost is $\text{\euro{}}47.1~$M and the minimum preventive investment that would have prevented or mitigated the scenario is  $\text{\euro{}}10.8~$M, thus $D_{\text{econ},S3}^{\textit{\text{und}}}=0.987$.

\paragraph{Regulatory sub-index}
$D_{\text{reg},3}^{\textit{\text{vul}}}$ is computed over $N_{v}^{\text{ref}} = 10$ control categories defined by IEC~62443-2-1:2024~\cite{IEC62443} and NERC~CIP (CIP-002–014)~\cite{NERCCIP}.

Scenario~3 assumes full compromise across all ten categories, as reported in Table~\ref{tab:reg_controls}, reflecting a state-sponsored attack with the capability and intent to exploit all regulatory control layers simultaneously (i.e., worst-case scenario) to maximize attack impacts. This yields $D_{\mathrm{reg},3}^{\mathrm{vul}} = 1.000$.

\begin{table}[t]
\footnotesize
\centering
\caption{Regulatory control categories and compromise status (Scenario 3).}
\begin{tabular}{lcc}
\toprule
{Category} & {Basis} & {Status} \\
\midrule
Access control & IEC/NERC & Compromised \\
Patch management & IEC/NERC & Compromised \\
Default credentials & IEC/NERC & Compromised \\
Network segmentation & IEC/NERC & Compromised \\
DER cybersecurity compliance & IEC/NERC & Compromised \\
OT monitoring & IEC/NERC & Compromised \\
Incident response (CIP-008) & NERC & Compromised \\
Supply chain risk (CIP-013) & NERC & Compromised \\

Physical security (CIP-006) & NERC & Compromised \\
Training and awareness (CIP-004) & NERC & Compromised \\
\bottomrule
\end{tabular}
\label{tab:reg_controls}
\end{table}

Adopting the weights $w_e = w_r = 0.35$ and $w_c = 0.30$ defined in Section~\ref{ss:EcReg}, the economic-regulatory sub-index for Scenario 3 yields $D_{\mathrm{econ\text{-}reg},S3} = 0.992$.

\subsubsection{Endogenous Core and Robustness to the Coupling Parameter}
\label{ss:Robust}

The endogenous core $\mathcal{M}(S_i;\gamma_i)$ is evaluated for each scenario using Eq.~\eqref{eq:10} and the index values in Table~\ref{tab:end_index_vals}. Scenario~1 is assigned to the additive regime ($\gamma_1 = 0$) and Scenarios~2 and~3 to the coupled regime ($\gamma_2 = \gamma_3 = 1$).
 
\begin{table}[t]
\footnotesize
\caption{Endogenous index values per scenario.}
\label{tab:end_index_vals}
\centering
\begin{tabular}{lccc}
\toprule
Quantity & $S_1$ & $S_2$ & $S_3$ \\
\midrule
$D_{\mathrm{phy},i}$ & 0.127 & 0.743 & 1.000 \\
$D_{\mathrm{op},i}$  & 0.214 & 0.705 & 0.799 \\
$D_{\mathrm{cyb},i}$ & 0.056 & 0.667 & 1.000 \\
$\bar{D}_i$          & 0.132 & 0.705 & 0.933 \\
$\Pi_i = \prod_{k} D_{k,i}$ & 0.0015 & 0.349 & 0.799 \\
\midrule
Assigned regime & additive & coupled & coupled \\
$\gamma_i$      & $0$      & $1$ & $1$ \\
\bottomrule
\end{tabular}
\end{table}
 
The regime assignments follow directly from the index values, in $S_1$, $D_{\mathrm{cyb},1} = 0.056$ yields $\Pi_1 = 0.0015$, rendering the coupling term negligible and confirming the additive assignment. In $S_2$ and $S_3$, all sub-indices are jointly large, yielding $\Pi_i \in \{0.349,\,0.799\}$; the coupling term contributes $33\%$ and $46\%$ of $\mathcal{M}$, respectively, a degradation component the additive term alone cannot reproduce.
 
Table~\ref{tab:12} numerically verifies the ranking-invariance property stated in Eq.~\eqref{eq:10}; although $\mathcal{M}$ and the $\mathcal{MDRI}$ ratios scale with $\gamma_i$, the ordering $\mathcal{MDRI}_3 > \mathcal{MDRI}_2 > \mathcal{MDRI}_1$ is preserved for all tested values, confirming that all qualitative conclusions of the framework are robust to the specific value of $\gamma_i$.
 
\begin{table}[t]
\centering
\caption{Numerical verification of $\mathcal{MDRI}$ ranking invariance to $\gamma_i$. $\mathcal{M}(S_1;0) = 0.132$ and is invariant to $\gamma_i$ by regime assignment.}
\label{tab:12}
\resizebox{0.48\textwidth}{!}{%
\begin{tabular}{cccccc}
\toprule
$\gamma_i$ & $\mathcal{M}(S_2;\gamma_i)$ & $\mathcal{M}(S_3;\gamma_i)$ &
$\mathcal{MDRI}_2/\mathcal{MDRI}_1$ & $\mathcal{MDRI}_3/\mathcal{MDRI}_2$ &
$\mathcal{MDRI}_3/\mathcal{MDRI}_1$ \\
\midrule
0.00 & 0.705 & 0.933 &  5.33 & 4.65 &  24.79 \\
0.25 & 0.792 & 1.133 &  5.99 & 5.03 &  30.10 \\
0.50 & 0.880 & 1.333 &  6.65 & 5.33 &  35.40 \\
1.00 & 1.054 & 1.732 &  7.97 & 5.78 &  46.02 \\
2.00 & 1.404 & 2.531 & 10.61 & 6.34 &  67.24 \\
5.00 & 2.452 & 4.928 & 18.53 & 7.07 & 130.93 \\
\bottomrule
\end{tabular}%
}
\end{table}

\subsubsection{Multidimensional Resilience Index}

The $\mathcal{MDRI}$ is evaluated via Eq.~\eqref{eq:11}, exogenous amplifiers apply exclusively to Scenario~3, which is the only scenario incorporating climatic and economic-regulatory stressors; for Scenarios~1 and~2, $D_{j,i} = 0$ for all $j \in \mathcal{K}_{\mathrm{ext}}$, and the $\mathcal{MDRI}$ reduces to the endogenous core. Table~\ref{tab:13} reports the complete breakdown.

\begin{table}[t]
\footnotesize
    \caption{Final $\mathcal{MDRI}$ values per scenario.}
    \label{tab:13}
    \centering
    \begin{tabular}{lccc}
        \toprule
        Component & $S_1$ & $S_2$ & $S_3$ \\
        \midrule
        $\mathcal{M}(S_i;\gamma_i)$            & 0.132 & 1.054 & 1.732 \\
        $(1 + D_{\mathrm{clim},i})$            & ---   & ---   & 1.765 \\
        $(1 + D_{\mathrm{econ\text{-}reg},i})$ & ---   & ---   & 1.992 \\
        $\mathcal{MDRI}_i$                     & 0.132 & 1.054 & 6.090 \\
        \bottomrule
    \end{tabular}
\end{table}

\paragraph{Diagnostic decomposition across scenario transitions}
Because Eq.~\eqref{eq:11} expresses $\mathcal{MDRI}$ as the product of an endogenous core and exogenous amplifiers, changes across a transition $S_i \to S_j$ decompose 
under a logarithmic transformation into additive contributions:
\[
\ln\!\left(\frac{\mathcal{MDRI}_j}{\mathcal{MDRI}_i}\right)
= \ln\!\left(\frac{\mathcal{M}_j}{\mathcal{M}_i}\right)
+ \sum_{k \in \mathcal{K}_{\mathrm{ext}}}
  \ln\!\left(\frac{1 + D_{k,j}}{1 + D_{k,i}}\right),
\]
allowing the relative weight of each mechanism to be expressed as a percentage. Table~\ref{tab:14} reports the results for the three scenario transitions.

\begin{table}[t]
\footnotesize
    \caption{Decomposition of $\mathcal{MDRI}$ amplification across scenario transitions.}
    \label{tab:14}
    \centering
    \begin{tabular}{lccc}
        \toprule
        Transition & $S_1 \to S_2$ & $S_2 \to S_3$ & $S_1 \to S_3$ \\
        \midrule
        $\mathcal{MDRI}$ ratio
            & 7.97  & 5.78  & 46.02 \\
        $\ln(\mathcal{MDRI}_j/\mathcal{MDRI}_i)$
            & 2.08  & 1.75  & 3.83  \\
        \midrule
        Endogenous $\ln(\mathcal{M}_j/\mathcal{M}_i)$
            & 2.08\,(100\%) & 0.50\,(28\%) & 2.57\,(67\%) \\
        Exogenous $\displaystyle\sum_{k}\ln\!\left(\frac{1+D_{k,j}}{1+D_{k,i}}\right)$
            & 0.00\,(0\%)   & 1.26\,(72\%) & 1.26\,(33\%) \\
        \bottomrule
    \end{tabular}
\end{table}

The decomposition reveals three distinct regimes:

\textit{(i)~Endogenously-driven degradation ($S_1 \to S_2$):} the transition from a single-vector to a multi-vector cascading attack yields an $\mathcal{MDRI}$ ratio of $7.97$, attributable entirely to the activation of cross-dimensional coupling, demonstrating that simultaneous compromise of physical, operational, and digital-cyber dimensions produces a degradation nearly eight times greater than an isolated attack.

\textit{(ii)~Exogenously-driven amplification ($S_2 \to S_3$):} the introduction of extreme climatic and economic-regulatory stressors yields a ratio of $5.78$, with $72\%$ of amplification driven by exogenous factors, reflecting a genuine shift in the dominant source of degradation rather than an artifact of endogenous saturation. 

\textit{(iii)~Compound degradation ($S_1 \to S_3$):} the combined effect produces a ratio of $46.02$, with $67\%$ driven by endogenous coupling and $33\%$ by exogenous amplification, revealing a sequential escalation in which the system first evolves within an endogenous regime 
($\mathcal{M}$ increasing from $0.132$ to $1.732$) before exogenous amplifiers intensify the resulting degradation.


\section{Research Gaps and Future Research Directions}
\label{s:Futur}

The synthesis and case study developed in Sections~\ref{s:syst} and~\ref{s:Comp} reveal a recurring limitation in current resilience assessment: EPS' resilience is largely treated as a set of separable attributes evaluated under nominal coordination assumptions. 

\subsection{Structural Gaps in Multidimensional Resilience Assessment}
\label{subsec:gaps_structural}

Three interconnected gaps emerge from the patterns identified in Sections~\ref{s:syst} and~\ref{s:Comp}. First, \textit{endogenous 
co-degradation} remains insufficiently modeled as a closed-loop process, as physical, operational, and digital-cyber dimensions are commonly treated as parallel rather than mutually constitutive. The case study demonstrates the consequences of this simplification: transitioning from a single to a 
multi-vector attack ($S_1\rightarrow S_2$) increases the $\mathcal{MDRI}$ ratio to $7.97$ solely through endogenous coupling, confirming that additive approaches systematically underestimate nonlinear cross-dimensional degradation.

Second, \textit{climatic and economic-regulatory dimensions} are often evaluated under static assumptions that isolate acute disturbances from chronic stress accumulation and presume intact supporting infrastructure, neglecting interactions with endogenous system dynamics. The case study attributes $72\%$ of the additional degradation in $S_2\rightarrow S_3$ to these exogenous amplifiers, while the limited cyber--climatic and cyber--economic literature in Figure~\ref{fig:4}(b) further 
reflects this gap. Third, \textit{coordination infrastructure} is commonly assumed fully operational during severe disruptions, even though a cyberattack disconnecting $74\%$ of inverter-based capacity would likely compromise the same communication and 
observability channels required for coordinated restoration. Addressing this requires modeling coordination infrastructure as a dynamic component subject to degradation, partial observability, and constrained recovery, rather than as an implicit boundary 
condition.

\subsection{Analytical Scope of the $\mathcal{MDRI}$ Framework}
\label{subsec:gaps_mdri}
The $\mathcal{MDRI}$ framework directly addresses the three gaps identified in Section~\ref{subsec:gaps_structural}. By separating additive and coupled degradation regimes through $\gamma_i$, it distinguishes endogenous core degradation from exogenous amplification and satisfies a ranking-invariance property across coupling strengths, providing analytical rigor beyond proxy-based aggregations. The framework has two applicability boundaries that delimit the scope of the present contribution.

First, the coupling parameter $\gamma_i$ is formulated as a structural rather than empirically calibrated quantity; as shown in Section~\ref{ss:Robust}, the monotone ordering of both $\bar{D}_i$ and $\Pi_i$ across scenarios guarantees that the $\mathcal{MDRI}$ ranking is invariant for any $\gamma_i \geq 0$, so that the qualitative conclusions of the framework do not depend on its specific value. 

Second, the modified IEEE~39-bus cyber-physical system is intended to demonstrate scenario discrimination and ranking robustness rather than broad generalizability across network scales and disturbance typologies.
These boundaries do not invalidate the framework but define the conditions under which its extension requires further empirical grounding.

\subsection{Emerging Challenges and Research Directions}
\label{subsec:gaps_future}

Energy-system transitions are reshaping the conditions for resilience assessment. The replacement of synchronous generation with inverter-based resources is particularly critical: the case study exhibits dynamic collapse due to PV penetration, while fragility thresholds derived for inertia-dominated systems become increasingly inadequate under 60--80\% renewable penetration. Simultaneously, the electrification of transport, heating, and industrial loads, climate non-stationarity, and regulatory fragmentation introduce coupled physical, cyber, climatic, and institutional disturbances, defining a research agenda across three horizons.

\paragraph{Validation and scalability} The immediate priority is extending the framework to larger heterogeneous benchmarks, including IEEE~118-bus, 300-bus, and synthetic large-scale systems under diverse disturbance scenarios, to test $\mathcal{MDRI}$ ranking invariance and 
establish empirical baselines for cross-study comparison.

\paragraph{Framework refinements}
Three main extensions are envisioned: \emph{i)}~a phase-resolved $\mathcal{MDRI}$ based on time-dependent sub-indices $D_{k,i}(t)$, enabling phase-sensitive dynamic monitoring instead of post-event diagnosis; \emph{ii)}~data-driven estimation methods for real-time assessment, allowing $\mathcal{MDRI}$ to operate as an observable resilience layer; and \emph{iii)}~integration of $\mathcal{MDRI}$ into transmission expansion and infrastructure hardening models as an optimization objective or constraint, elevating it from a descriptive metric to a prescriptive design criterion.

\paragraph{Paradigm-level extensions}
Beyond EPS, $\mathcal{MDRI}$ can be extended to interdependent infrastructure planning, including coupled electricity, transport, water, and telecommunications networks exposed to cascading failures. In this context, the framework could support coordinated resilience optimization across interconnected critical infrastructures.

\section{Conclusions}
\label{s:Con}

This study demonstrates that resilience in modern EPS cannot be captured through isolated dimensional assessments; rather, it emerges from the interaction of endogenous and exogenous factors under compound stress conditions.
The systematic review reveals a fragmented research landscape: 81\% of existing studies address no more than two resilience dimensions, none integrates all five, and critical coupling pathways, particularly cyber-climatic and cyber-economic interactions, remain largely unexplored. The proposed $\mathcal{MDRI}$ addresses this gap by simultaneously capturing endogenous coupling effects and exogenous amplification mechanisms across all five dimensions.
Validation under three escalating cyber-physical scenarios inspired by the coordinated multi-vector attack on Polish energy infrastructure in December 2025 provides clear quantitative evidence. A single-vector attack produces a resilience loss index of 0.132, whereas a cascading multi-vector attack across the three endogenous dimensions increases it to 1.054, a 7.97-fold rise driven entirely by cross-dimensional coupling, culminating in dynamic collapse under compounded endogenous stress.
 When concurrent climatic and economic-regulatory stressors are incorporated, the index rises further to 6.090, representing a 46-fold increase over baseline conditions, with exogenous amplification becoming the dominant degradation mechanism. Assessments confined to a single dimension are structurally incapable of detecting this progression.
These findings carry direct implications for resilience planning. Cross-dimensional coupling dominates the early stages of cascading disruption, whereas exogenous stressors become the primary amplification drivers under fully compounded conditions. Strengthening individual dimensions without addressing their coupling pathways may therefore reduce local fragility while leaving systemic vulnerability largely intact.
Although validated on a benchmark system with a structurally defined coupling parameter, the $\mathcal{MDRI}$ establishes a quantitative foundation for future extensions to large-scale heterogeneous networks and cross-infrastructure resilience modeling.
Ultimately, compound resilience assessment is not an incremental extension of existing methodologies but a fundamentally different analytical problem. The $\mathcal{MDRI}$ reframes resilience from a specific attribute into an emergent cross-dimensional property of modern power systems, one that becomes visible and measurable only when all dimensions are assessed together.

\section{Acknowledgment}
This work is funded in part by the National Science Foundation (NSF) Award Number \#2501975.
 
\section{Declaration of generative AI and AI-assisted technologies in the manuscript preparation process}

During the preparation of this work, the authors used AI-assisted literature discovery tools (e.g., Consensus and Rabbit) to enhance the efficiency of  scientific literature discovery. These tools did not replace critical evaluation, interpretation, or scientific judgment. The authors reviewed, verified, and edited all content and take full responsibility for the published article.

\newpage
\appendix
\clearpage 
\section{Table Appendix Section}\label{app1}


{\scriptsize
\onecolumn
\begin{longtable}{||p{0.07\textwidth}|p{0.04\textwidth}|p{0.13\textwidth}|p{0.15\textwidth}|p{0.10\textwidth}|p{0.09\textwidth}|p{0.17\textwidth}|p{0.03\textwidth}||}

\caption{Taxonomy of Metrics and Methods for Physical Resilience.}
\label{tab:2}\\

\hline \hline
\textbf{Study area} &
\textbf{Scope} &
\textbf{Disturbance} &
\textbf{Formulation} &
\textbf{Metric} &
\textbf{Mitigation} &
\textbf{Limitation} &
\textbf{Ref.}\\
\hline \hline
\endfirsthead

\hline \hline
\textbf{Study area} &
\textbf{Scope} &
\textbf{Disturbance} &
\textbf{Formulation} &
\textbf{Metric} &
\textbf{Mitigation} &
\textbf{Limitation} &
\textbf{Ref.}\\
\hline \hline
\endhead

\hline
\multicolumn{8}{r}{\textit{}}\\
\endfoot

\hline \hline
\endlastfoot

\multirow{3}{=}{\raggedright Distribution planning \& resource allocation} &
Phys. &
Renewable, load, and operational variability, $N-1$ contingencies &
Multi-objective AC/DC OPF, metaheuristic and stochastic optimization, evolutionary algorithms &
Power losses, voltage margins, reliability indices, hosting capacity, planning cost &
Infrastructure sizing and placement, DER allocation &
Single-dimension scope, physical damage treated as a static boundary condition masking co-degradation with operational and digital-cyber layers &
\cite{Islam2025Optimizing, en15186682, en16165907, 10373923,
      JEON2025114793, GUPTA2023120955, 10638037}\\ \cline{2-8}

&
Phys.--Op. &
Network operational constraints, load variability, outage recovery requirements &
Distributed OPF, deep reinforcement learning for reactive power and network scheduling, joint topology and dispatch optimization &
Power balance, DER utilization efficiency, switching cost, restoration performance &
Coordinated network reconfiguration, DER scheduling &
Intact digital infrastructure assumptions, operational decisions modeled independently of cyber degradation blocking detection of endogenous coupling &
\cite{10638037, PANG2024114068, ZHAO2025101720, en18092196,
      ALVAREZ2024101413}\\ \cline{2-8}

&
Phys.--Eco. &
Market price volatility, renewable variability, investment risk exposure &
Bilevel and multi-level expansion models, investment planning under market uncertainty, and stochastic optimization &
Investment costs, system and capacity adequacy, market power &
Economic planning of network reinforcement and DER expansion &
Economic signals evaluated under intact infrastructure, decouples investment decisions from physical damage dynamics under HILP conditions &
\cite{10965700, Akbari2025, su14052998, CHO2022107924}\\ \hline

Long-term expansion planning &
Phys.--Op.--Eco. &
Demand growth projections, renewable, market, and equipment variability, climate-driven generation uncertainty &
MILP/MINLP expansion planning, multi-level and multi-period planning models, risk-averse stochastic optimization &
Investment and operational cost, system adequacy, capacity expansion pathways &
Long-term infrastructure investment, generation and transmission expansion &
Broadest physical scope reviewed, yet aggregated temporal resolution prevents modeling of cascading cross-dimensional failures, understating compound vulnerability &
\cite{HAMIDPOUR2022122321, Rajakumar2025, XU2022102438,
      PANG2024114068}\\ \hline

\multirow{2}{=}{\raggedright Asset monitoring \& lifecycle management} &
Phys. &
Operational stress, equipment aging, environmental exposure, historical degradation patterns &
Statistical and machine learning degradation models, graph learning for health assessment &
Asset health index, failure probability, remaining useful life, lifecycle risk &
Component-level condition monitoring, lifecycle management &
Component-level scope, failure propagation across interconnected physical and cyber layers remains unmodeled &
\cite{en18051254, Rajora_Sanz-Bobi_Domingo_2022, 10926503}\\ \cline{2-8}

&
Phys.--Cyb. &
Sensor variability, communication-dependent monitoring, cyber-induced data loss, equipment aging due to digitization &
Digital-twin frameworks, deep and graph learning, data-driven degradation monitoring integrating SCADA/IoT data streams &
Degradation rate, remaining useful life, failure probability, anomaly detection accuracy &
Predictive maintenance, digital-twin-enabled condition monitoring &
Continuous sensor availability assumptions, degraded cyber observability under simultaneous physical stress remains unmodeled, suppressing detection of endogenous coupling &
\cite{10945887, ElRashidy2025TransformerHealth,
      Olojede2025PowerGridML, Sankaran2025BigDataPredictiveMaintenance}\\

\end{longtable}
\twocolumn
}

{\scriptsize
\onecolumn
\begin{longtable}{||p{0.07\textwidth}|p{0.06\textwidth}|p{0.11\textwidth}|p{0.13\textwidth}|p{0.09\textwidth}|p{0.11\textwidth}|p{0.16\textwidth}|p{0.03\textwidth}||}

\caption{Taxonomy of Metrics and Methods for Operational Resilience.}
\label{tab:3}\\

\hline \hline
\textbf{Study area} &
\textbf{Scope} &
\textbf{Disturbance} &
\textbf{Formulation} &
\textbf{Metric} &
\textbf{Mitigation} &
\textbf{Limitation} &
\textbf{Ref.}\\
\hline \hline
\endfirsthead

\hline \hline
\textbf{Study area} &
\textbf{Scope} &
\textbf{Disturbance} &
\textbf{Formulation} &
\textbf{Metric} &
\textbf{Mitigation} &
\textbf{Limitation} &
\textbf{Ref.}\\
\hline \hline
\endhead

\hline
\multicolumn{8}{r}{\textit{}}\\
\endfoot

\hline \hline
\endlastfoot

\multirow{2}{=}{\raggedright Post-fault reconfiguration and restoration} &
Op.--Phys. &
Line faults, feeder outages, weather-induced failures &
Security-constrained reconfiguration, scenario-based and
topology-constrained stochastic optimization, linear programming &
Power balance, voltage security, load shedding, restoration
performance, switching cost & 
Infrastructure sizing and siting, DER allocation &
Operational decisions modeled without cyber-layer feedback, coordination assumed intact precisely when digital degradation would suppress it, blocking endogenous coupling detection &
\cite{SHI2021106355, en18226062, en16155740, en18020266}\\ \cline{2-8}

&
Op.--Phys.--Clim. &
Extreme precipitation, hurricane damage, weather-induced
infrastructure failures &
Security-constrained dispatch under hazard scenarios, stochastic
optimization with climate-driven contingency sets &
Power balance, voltage security, load shedding under
hazard-constrained operation &
Hazard-constrained reconfiguration, resilience-driven
restoration under extreme events &
Climatic stress modeled as exogenous hazard only, interaction with communication degradation and operational decision quality under compound conditions unaddressed &
\cite{WANG2025111214, asi8050149}\\ \hline

Security-constrained dispatch and unit commitment &
Op.--Phys. &
Equipment outages, demand and renewable variability,
$N-1$ contingencies &
Security-constrained economic dispatch, unit commitment,
deterministic and stochastic optimization, scenario-based scheduling &
Operating cost, load shedding, reserve margins,
voltage security, power balance &
System-level generation scheduling, operational dispatch &
Contingency sets fixed ex-ante, progressive co-degradation of physical assets and digital situational awareness under evolving disruptions remains unmodeled &
\cite{en18205454, en15072505, brun2025alternating, 11016705}\\ \hline

\multirow{4}{=}{\raggedright Service restoration strategies} &
Op.--Phys. &
Post-disaster outages, infrastructure repair requirements, and
renewable and load variability &
Metaheuristic and stochastic optimization, hybrid restoration
sequencing, network and load prioritization constraints &
Restoration time, unserved energy, load prioritization,
switching operations, service continuity &
Coordinated microgrid recovery sequential service restoration &
Centralized coordination assumed available, cyber-layer degradation that fragments situational awareness under severe physical damage excluded from restoration feasibility &
\cite{Hamadneh2025, en18184833, engproc2022020038, en18174620}\\ \cline{2-8}

&
Op.--Eco. &
Post-disaster energy supply gaps, shared resource allocation
under market constraints &
Co-optimization of energy scheduling and storage sharing,
multi-body game-theoretic frameworks, stochastic dispatch &
Operational cost, energy balance, storage utilization,
service continuity &
Market-based coordinated recovery, shared energy storage
management &
Market mechanisms evaluated under intact infrastructure, economic-physical feedback under disruption excluded, masking how cost-optimal decisions amplify rather than contain physical damage &
\cite{BABAEI2024110167, s25020406}\\ \cline{2-8}

&
Op.--Phys.--Clim. &
Post-disaster outages under compound climatic stressors,
weather-driven infrastructure repair requirements &
Distributionally robust optimization, stochastic restoration
under multiple uncertainties including climatic variability &
Restoration time, unserved energy, load shedding under
compound hazard conditions &
Climate-aware coordinated restoration &
Climatic stress modeled on nominally intact system, simultaneous physical and digital-cyber degradation under compound stressors excluded, understating exogenous amplification &
\cite{ZHANG2024110367}\\ \cline{2-8}

&
Op.--Cyb. &
Communication-dependent recovery, multi-agent coordination
under partial digital infrastructure availability &
Distributed multi-agent energy management, co-simulation
frameworks integrating cyber and operational layers &
Service continuity, coordination efficiency,
communication-dependent restoration performance &
Cyber-aware distributed restoration coordination &
Cyber failures modeled as bounded and independent, closed-loop feedback between communication loss and operational decision quality, the core endogenous coupling mechanism, remains unrepresented &
\cite{en18174620}\\ \hline

\multirow{3}{=}{\raggedright Coordinated DER operation} &
Op.--Phys. &
Load variability, renewable intermittency,
network operational constraints &
DER scheduling, distributed OPF, microgrid energy management, and
decomposition-based multi-objective optimization &
Operational cost, load shedding, power balance security,
DER utilization efficiency &
Coordinated DER dispatch, multi-area microgrid operation &
Synchronized digital control assumed, absence of cyber-layer modeling prevents capture of coordination collapse under simultaneous physical and communication disruption &
\cite{10385080, ZHANG2022108197, 10374172, 10833641}\\ \cline{2-8}

&
Op.--Eco. &
Load variability, renewable intermittency,
market-driven DER participation &
Multi-objective scheduling with cost and emission objectives,
quantum-inspired optimization, market-integrated DER management &
Operational cost, emission reduction,
DER utilization efficiency, market participation &
Cost- and emission-aware DER coordination &
Economic optimization decoupled from physical disruption dynamics, market-driven DER coordination cannot detect how infrastructure degradation renders cost-optimal dispatch infeasible under compound stress &
\cite{KUMAR2024109702, Paul2025}\\ \cline{2-8}

&
Op.--Phys.--Cyb. &
Cyber-induced disruptions affecting DER coordination,
network reconfiguration under compromised digital infrastructure &
Hardware-in-the-loop validation, integrated reconfiguration
and scheduling under cyber-physical constraints &
Restoration performance, coordination efficiency
under partially degraded digital infrastructure &
Cyber-resilient DER coordination and microgrid management &
Widest scope in this paradigm yet validated under partial failure only, complete co-degradation across all three endogenous dimensions, ,the condition under which cross-dimensional amplification emerges, remains uncharacterized &
\cite{10833641}\\

\end{longtable}
}


{\scriptsize
\onecolumn
\begin{longtable}{||p{0.06\textwidth}|p{0.07\textwidth}|p{0.11\textwidth}|p{0.13\textwidth}|p{0.09\textwidth}|p{0.11\textwidth}|p{0.16\textwidth}|p{0.03\textwidth}||}

\caption{Taxonomy of Metrics and Methods for Digital-Cyber Resilience.}
\label{tab:4}\\

\hline \hline
\textbf{Study area} &
\textbf{Scope} &
\textbf{Disturbance} &
\textbf{Formulation} &
\textbf{Metric} &
\textbf{Mitigation} &
\textbf{Limitation} &
\textbf{Ref.}\\
\hline \hline
\endfirsthead

\hline \hline
\textbf{Study area} &
\textbf{Scope} &
\textbf{Disturbance} &
\textbf{Formulation} &
\textbf{Metric} &
\textbf{Mitigation} &
\textbf{Limitation} &
\textbf{Ref.}\\
\hline \hline
\endhead

\hline
\multicolumn{8}{r}{\textit{}}\\
\endfoot

\hline \hline
\endlastfoot

\multirow{3}{=}{\raggedright Observability \& state estimation} &
Cyb. &
Measurement noise, partial observability,
false data injection &
Static and dynamic state estimation, centralized
PMU-SCADA fusion, probabilistic estimators &
Estimation accuracy, observability index,
detection rate &
Wide-area monitoring, PMU-SCADA data fusion &
Single-dimension scope, physical damage as degrader of cyber observability unmodeled, severing the feedback loop through which reduced situational awareness amplifies physical collapse &
\cite{DAI2024110203}\\ \cline{2-8}

&
Cyb.--Op. &
Communication failures, renewable variability,
degraded real-time operational visibility &
Physics-informed and graph neural network estimators,
distributed PMU-SCADA fusion, AI-enhanced
situational awareness &
Estimation accuracy, observability index,
situational awareness quality &
Wide-area monitoring, AI-driven operational
awareness under partial observability &
Failures modeled as independent stochastic events, simultaneous physical-cyber co-degradation, the condition under which observability loss accelerates operational collapse—excluded &
\cite{NGO2024122602, KOTHA2022107794, Zhu2021AIGrid, pr11051509}\\ \cline{2-8}

&
Cyb.--Phys. &
Physical network changes affecting measurement
topology, PMU placement under infrastructure constraints &
Integer linear programming for optimal PMU placement
considering network topology &
Observability index, network coverage
under physical infrastructure constraints &
Infrastructure-aware PMU deployment &
PMU placement is optimized for static topology, dynamic physical degradation that invalidates observability coverage and with it, operational coordination, remains unmodeled &
\cite{en17092140}\\ \hline

\multirow{2}{=}{\raggedright Communication networks \& ICT infrastructure} &
Cyb. &
Network congestion, protocol incompatibilities,
legacy interoperability constraints, SCADA architectures &
Review and classification of smart grid communication
architectures, SCADA system surveys, IT-OT integration
frameworks &
Communication standards compliance,
architectural interoperability &
Smart grid communication network design,
IT-OT integrated systems &
Architectural characterization under nominal conditions only, cross-layer failure propagation between communication infrastructure and physical assets excluded, masking endogenous coupling pathways &
\cite{9759422, Vedantham2022ICTSLES, 9907002}\\ \cline{2-8}

&
Cyb.--Op.&
Communication latency, delayed control responses,
packet errors affecting microgrid coordination &
Time-delay system modeling, delay-aware secondary
control, communication-constrained distributed
frequency regulation &
Communication latency, packet loss rate,
control stability margins, frequency deviation &
Delay-tolerant control architectures,
latency-aware distributed coordination &
Latency modeled as isolated impairment, correlated cyber-physical outages that simultaneously degrade control performance and physical stability margins remain unrepresented &
\cite{GAO2026112639, en18061361, 10990218}\\ \hline

\multirow{3}{=}{\raggedright Cybersecurity \& attack mitigation} &
Cyb. &
False data injection, DoS attacks, anomaly detection
in industrial sensor networks and SCADA systems &
Supervised and unsupervised learning, anomaly detection,
feature selection, time-series analysis &
Detection accuracy, false positive
and negative rate &
Data-driven intrusion detection,
SCADA/ICS security &
Attack detection trained on stable infrastructure, physical degradation that shifts detection model distributions, enabling undetected cascading into operational collapse—excluded &
\cite{KAYODESAHEED2023101532}\\ \cline{2-8}

&
Cyb.--Phys. &
Firmware attacks on microgrids, DER cybersecurity
vulnerabilities, coordinated attacks propagating
into the physical layer &
Cyber-physical co-simulation, time-series attack
detection, vulnerability modeling for inverter-based
and DER-connected systems &
Detection accuracy, physical impact
quantification, vulnerability exposure index &
Device-level vulnerability modeling,
DER-focused attack analysis, firmware security &
Attacks modeled as static and contained, feedback through which physical damage degrades cyber detection capacity—enabling further undetected compromise—remains unmodeled as a closed-loop process &
\cite{zografopoulos2022time, zografoopulos2023distributed,
      9633019, zografopoulos2021cyber}\\ \cline{2-8}

&
Cyb--Phys.--Op. &
Coordinated cyber-physical attacks affecting both
operational decisions and physical infrastructure integrity &
Cyber-physical interdependence modeling, threat
assessment integrating operational and physical
impact layers &
Cyber-physical risk metrics, operational
impact of cyber incidents, cascading failure
propagation across layers &
Integrated cyber-physical security assessment &
Widest scope in this paradigm yet attacks treated as predictable and contained, simultaneous co-degradation across all three endogenous dimensions, the mechanism producing nonlinear amplification, remains uncharacterized &
\cite{XU2021111642, zografopoulos2025cyber}\\ \hline

\multirow{2}{=}{\raggedright Distributed \& intelligent digital control} &
Cyb.--Op. &
Renewable and load variability, operational stress
under partially degraded communication &
Hierarchical and multi-agent control, event-triggered
communication, multi-agent deep reinforcement learning
for distributed real-time control &
Frequency and voltage deviation,
stability margins, coordination efficiency &
Hierarchical and distributed control,
real-time intelligent control for AC/DC microgrids &
Validated under bounded delays only, collapse of coordination guarantees under simultaneous physical damage and communication loss, where endogenous coupling is most critical and unaddressed &
\cite{su17125423, MBUNGU2025124486, en16052445,
      Zhen2025MultiAgentDRL}\\ \cline{2-8}

&
Cyb--Phys.--Op. &
Physical network constraints affecting distributed
control, hybrid energy storage under cyber-physical
operational stress &
OPF and state-space stability models, hierarchical
control integrating physical network and cyber
communication constraints, adaptive event-triggered
coordination &
Frequency and voltage deviation, stability margins
under physical and communication constraints &
Cyber-physically integrated hierarchical control,
adaptive coordination under joint physical-cyber stress &
Validated under partial disruption only, feedback between physical damage, sensor loss, and control instability under compound conditions remains uncharacterized &
\cite{9394790, electronics14163303}\\

\end{longtable}
\twocolumn
}


{\scriptsize
\onecolumn
\begin{longtable}{||p{0.05\textwidth}|p{0.04\textwidth}|p{0.11\textwidth}|p{0.15\textwidth}|p{0.09\textwidth}|p{0.08\textwidth}|p{0.15\textwidth}|p{0.03\textwidth}||}
\caption{Taxonomy of Metrics and Methods for Climatic-External Resilience.}

\label{tab:5}\\

\hline \hline
\textbf{Study area} &
\textbf{Scope} &
\textbf{Disturbance} &
\textbf{Formulation} &
\textbf{Metric} &
\textbf{Mitigation} &
\textbf{Limitation} &
\textbf{Ref.}\\
\hline \hline
\endfirsthead

\hline \hline
\textbf{Study area} &
\textbf{Scope} &
\textbf{Disturbance} &
\textbf{Formulation} &
\textbf{Metric} &
\textbf{Mitigation} &
\textbf{Limitation} &
\textbf{Ref.}\\
\hline \hline
\endhead

\hline
\multicolumn{8}{r}{\textit{}}\\
\endfoot

\hline \hline
\endlastfoot

\multirow{3}{=}{\raggedright Extreme weather disruption} &
Clim.--Phys.&
Floods, windstorms, icing, heatwaves, wildfires,
low-temperature events, hazard intensity progression &
Fragility curves, reliability block diagrams,
Monte Carlo simulation, scenario-based vulnerability
assessment &
Fragility thresholds, failure probability,
performance degradation index,
generation loss &
Infrastructure hardening, transmission and
substation protection, DER exposure assessment &
Acute shock modeled on nominally intact system, chronic climatic stress that progressively erodes physical margins before the event excluded, understating exogenous amplification &
\cite{ZHANG2023109065, HAN2025115374,
      su152216044, karagiannakis_2025}\\ \cline{2-8}

&
Clim.--Phys.--Op.&
Floods, extreme precipitation, low-temperature events,
weather-driven transmission and distribution failures
requiring coordinated operational response &
Post-hazard OPF, MILP and stochastic programming,
mobile resource deployment under hazard constraints,
co-optimization of short- and long-term operational
decisions &
Energy not supplied, outage and restoration cost,
reliability indices, restoration performance
under hazard-constrained operation &
Hazard-constrained operation, restoration
resource allocation, mobile substation deployment &
Climatic stress modeled as standalone hazard, concurrent degradation of physical and digital-cyber dimensions that climatic events trigger remains unmodeled, masking exogenous amplification of endogenous coupling &
\cite{SOUTO2022107545, LI2024133677,
      SHUKLA2025101973, 11048914, 10422738}\\ \cline{2-8}

&
Clim.--Phys.--Op.--Eco. &
Flood-driven transmission failures with co-optimized
short-term operational and long-term investment decisions &
Bi-level co-optimization integrating hazard-constrained
operation with infrastructure investment planning,
stochastic programming across temporal scales &
Restoration cost, investment efficiency,
long-term resilience under recurring flood exposure &
Integrated operational and investment planning
under climate-driven flood risk &
Broadest scope in this paradigm yet fragility treated as static, cyber-layer degradation and bidirectional economic-operational feedback under acute climatic stress remain unaddressed, leaving the full compound mechanism unmodeled &
\cite{SHUKLA2025101973}\\ \hline

\multirow{4}{=}{\raggedright Climate-driven resource adequacy} &
Clim.--Phys. &
Chronic temperature rise, drought-driven reduction in
hydropower and thermoelectric generation capacity &
Long-term capacity adequacy assessment,
multi-scenario climate projections applied to
generation availability &
Annual and seasonal generation adequacy,
capacity shortfall under climate scenarios &
Generation adequacy planning under
chronic climatic stress &
Climate is treated as one-way exogenous forcing, feedback through which sustained stress ages infrastructure and reshapes fragility, the system that acute events then strike, is excluded &
\cite{SHUAI2024122977, ZHAO2023113141}\\ \cline{2-8}

&
Clim.--Phys.--Eco. &
Multi-decadal temperature and hydrological variability,
climate-driven generation cost and investment risk &
Multi-GCM ensemble projections coupled with
capacity expansion and investment planning models,
stochastic and multi-scenario optimization &
Annual energy balance, investment cost
under climate scenarios, capacity adequacy
pathways &
National and continental EPS planning
under long-term climate uncertainty &
Long-horizon planning under projected envelopes, short-term operational response and investment performance under tail climatic events that exceed the envelope remain unmodeled &
\cite{DEMARCO2025100235,
      MA2024120257}\\ \cline{2-8}

&
Clim.--Eco. &
Long-term temperature variability and its effect on
electricity investment returns, chronic climatic
stress under energy transition pathways &
Econometric and scenario-based investment modeling,
stochastic planning under climate-driven demand
and cost uncertainty &
Investment returns under climate scenarios,
capacity adequacy under transition constraints &
Climate-informed energy investment
and transition policy &
Economic projections decoupled from physical and operational feasibility, cyber-climatic and full endogenous pathways excluded, a scope reflecting the structural rarity of cyber-economic-climatic integration in the literature &
\cite{Khan2021TemperatureElectricity,
      BLOOMFIELD20211062}\\ \cline{2-8}

&
Clim.--Phys.--Op. &
Chronic climatic stress interacting with
generation dispatch and system-level adequacy &
Multi-model framework integrating short- and
long-term climatic influences on generation
dispatch and grid operation &
Seasonal energy balance, operational
adequacy under climatic variability,
generation dispatch performance &
Multi-scale climate-operational
adequacy assessment &
Climate modeled as one-way forcing, bidirectional feedback through which operational decisions reshape infrastructure resilience under sustained stress and vice versa excluded &
\cite{COHEN2022119193}\\ \hline

\multirow{4}{=}{\raggedright Compound climate hazards} &
Clim. &
Multi-risk meteorological events, compound wind
and cold co-occurrence affecting offshore systems &
Dynamic hidden Markov models for multi-risk
meteorological forecasting, statistical
characterization of compound hazard drivers &
Multi-risk forecasting accuracy,
compound event probability &
Statistical compound hazard assessment
and forecasting &
Pure hazard characterization without infrastructure interaction, system response under compound conditions outside scope, exemplifying the structural disconnect between climatic risk modeling and resilience assessment &
\cite{su17083606}\\ \cline{2-8}

&
Clim.--Phys.&
Concurrent compound extremes, heatwaves and power
outages, wind and rain loads on transmission towers,
natural hazard cascades on infrastructure &
Multi-hazard probabilistic risk assessment,
Kriging-based probabilistic framework,
joint-probability hazard modeling,
interdependent network cascade modeling &
Multi-hazard failure probability,
outage frequency and duration,
energy not served &
Surrogate fragility modeling,
joint-distribution inference,
infrastructure cascade assessment &
Multi-hazard modeling on nominally intact endogenous layers, operational and cyber co-degradation under compound climatic stressors, where exogenous amplification peaks—remains unrepresented &
\cite{Saki2025HeatwavesOutages, Rahman_2025,
      BI2023109615, MUHLHOFER2023109194,
      nhess-25-843-2025}\\ \cline{2-8}

&
Clim.--Econ. &
Renewable variability in long-term decarbonization
scenarios, compound climatic uncertainty affecting
investment pathways &
Scenario-based stochastic planning integrating
compound renewable variability with investment
and decarbonization objectives &
Decarbonization trajectory performance
under compound climatic variability,
investment adequacy &
Climate-aware long-term energy
transition planning &
Investment pathways modeled under fixed operational and physical conditions, compound climatic stress on the very infrastructure that investment strategies presuppose is excluded, masking economic-climatic-physical coupling &
\cite{FLORES2024123464}\\ \cline{2-8}

&
Clim.--Phys.--Op.&
Concurrent solar and wind variability stressing
integrated energy system sizing and dispatch &
Scenario-based stochastic optimization integrating
compound renewable variability into component
sizing and operational scheduling &
Energy balance, operational cost under
compound variability scenarios,
system sizing adequacy &
Integrated energy system design
under compound climatic variability &
Stationary stochastic climate inputs, non-stationary hazard dynamics and progressive infrastructure degradation under persistent multi-stressor conditions, the regime where exogenous amplification compounds with endogenous coupling—unmodeled &
\cite{HUA2024121543}\\

\end{longtable}
\twocolumn
}

{\scriptsize
\onecolumn
\begin{longtable}{||p{0.05\textwidth}|p{0.05\textwidth}|p{0.11\textwidth}|p{0.13\textwidth}|p{0.09\textwidth}|p{0.08\textwidth}|p{0.15\textwidth}|p{0.03\textwidth}||}
 
\caption{Taxonomy of Metrics and Methods for Economic-Regulatory Resilience.}
\label{tab:table6}\\
 
\hline \hline
\textbf{Study area} &
\textbf{Scope} &
\textbf{Disturbance} &
\textbf{Formulation} &
\textbf{Metric} &
\textbf{Mitigation} &
\textbf{Limitation} &
\textbf{Ref.}\\
\hline \hline
\endfirsthead
 
\hline \hline
\textbf{Study area} &
\textbf{Scope} &
\textbf{Disturbance} &
\textbf{Formulation} &
\textbf{Metric} &
\textbf{Mitigation} &
\textbf{Limitation} &
\textbf{Ref.}\\
\hline \hline
\endhead
 
\hline
\multicolumn{8}{r}{\textit{}}\\
\endfoot
 
\hline \hline
\endlastfoot

\multirow{4}{=}{\raggedright Investment incentives \& planning} &
Eco. &
Fuel price volatility, renewable variability,
market price uncertainty, investment risk exposure &
Risk-averse stochastic optimization, real options
analysis, capacity market design for generation
investment decisions &
Investment profitability, risk-adjusted returns,
cost recovery, capacity adequacy,
market power indicators &
Risk-based investment frameworks,
capacity market mechanisms &
Equilibrium models under normal conditions, price signals decoupled from physical and operational disruption, masking how economic optimization under HILP events amplifies rather than mitigates systemic fragility &
\cite{DIMANCHEV2024107639, LI2024109536,
      en16104241}\\ \cline{2-8}
 
&
Eco.--Op. &
Peak-shaving requirements, shared energy storage
allocation, market-driven ancillary service
compensation &
Co-evolution game-theoretic frameworks,
compensation mechanism design for storage
participation, stochastic dispatch with
market incentives &
Operational cost, storage utilization efficiency,
cost recovery, service continuity &
Market-based operational coordination,
storage-sharing incentive design &
Market-operational coordination evaluated under intact infrastructure, bidirectional feedback through which physical disruption renders market-driven dispatch infeasible, and degraded operations distort price signals are excluded &
\cite{ZHANG2025117127, Liang2025CapacityCompensation,
      systems13090817}\\ \cline{2-8}
 
&
Eco.--Phys. &
Infrastructure investment under adversarial
resource adequacy conditions, vulnerability-based
storage siting and sizing &
Adversarially robust adequacy estimation with
deep generative modeling, tri-level optimization
for storage siting under grid vulnerability &
Resource adequacy under adversarial conditions,
investment efficiency, vulnerability-weighted
capacity deployment &
Resilience-oriented infrastructure investment
and resource adequacy planning &
Investment optimization on static networks: bidirectional feedback between market-driven infrastructure decisions and dynamic damage evolution under cascading disruption remains unmodeled &
\cite{MASOUMI2025111374, ZHAO2025101720}\\ \cline{2-8}
 
&
Eco.--Phys.--Op. &
Transmission expansion under market competition,
security-constrained generation and transmission
co-optimization with bidding in energy and
reserve markets &
Bi-level stochastic optimization integrating
market bidding, security constraints, and
transmission expansion, scenario-based
planning with market power considerations &
Investment cost, market power, system adequacy,
operational security under market conditions &
Security-constrained expansion with
market participation &
Broadest scope in this paradigm yet markets modeled at equilibrium, cyber-layer absence combined with equilibrium clearing under physical damage leaves the full endogenous-exogenous interaction unrepresented &
\cite{DINI2021107017, en16073256}\\ \hline

\multirow{5}{=}{\raggedright Electricity market design} &
Eco. &
Renewable variability, demand-side price response,
strategic bidding and market power &
Dispatch-based market clearing, partial
equilibrium modeling of wholesale electricity
markets with intermittent power &
Wholesale price levels and volatility,
market clearing efficiency &
Wholesale market design, price
formation under intermittent supply &
Market clearing under intact infrastructure, the circular dependency through which normal-condition price signals systematically undervalue resilience &
\cite{MENDES2024107343, en18061435}\\ \cline{2-8}
 
&
Eco.--Op. &
Demand-side flexibility, ancillary service
requirements, EV and aggregator participation,
strategic bidding under renewable uncertainty &
Co-optimized energy and reserve clearing,
hierarchical and decomposition optimization,
multi-agent deep reinforcement learning for
wholesale market simulation &
Wholesale price, reserve adequacy,
flexibility value, aggregator revenue &
Co-optimized energy and ancillary markets,
aggregator participation frameworks &
Dispatch and bidding modeled under intact networks, agent behavior divergence triggered by extreme prices under physical degradation, where market clearing itself becomes unstable—excluded. &
\cite{LI2022119644, NIE2024109917, KROGER2023120406,
      HARDER2023100295, KOLTSAKLIS2023113084}\\ \cline{2-8}
 
&
Eco.--Cyb.&
Cyber-influenced market behavior, data-driven
multi-agent market simulation under information
asymmetry &
Multi-agent deep reinforcement learning for
wholesale market modeling under information
and communication constraints &
Wholesale price formation, agent learning
convergence, market efficiency under
information constraints &
Cyber-aware market simulation &
Sole reviewed study addressing the cyber-economic pair, cyberattacks treated as informational constraints rather than vectors propagating into operational and physical collapse &
\cite{HARDER2023100295}\\ \cline{2-8}
 
&
Eco.--Clim.&
Renewable intermittency as a climatic-driven
market constraint, seasonal and inter-annual
variability affecting market clearing &
Analytical market model with intermittent
renewable supply as an exogenous climatic input &
Price formation under intermittent supply,
market efficiency under renewable variability &
Market design adapted to high-renewable
penetration under climatic variability &
Climate treated as one-way market input, absence of endogenous dimensions prevents capture of compound climatic-economic stress that exogenous amplifiers are designed to model &
\cite{en18061435}\\ \cline{2-8}
 
&
Eco.--Phys.--Op.&
Transmission network constraints on market
clearing, demand-side participation and EV
aggregation under network limits &
Two-stage optimal market mechanism integrating
transmission and distribution network constraints
with aggregator and EV participation &
Wholesale price, network constraint costs,
aggregator revenue, demand flexibility value &
Network-constrained market design with
flexible demand participation &
Widest scope in this paradigm yet networks treated as static, absence of cyber layer combined with normal-condition market designs leaves the full mechanism through which infrastructure disruption invalidates clearing feasibility unmodeled &
\cite{NIE2024109917}\\ \hline

\multirow{3}{=}{\raggedright Regulatory governance \& policy frameworks} &
Eco. &
Carbon price volatility, permitting constraints,
regulatory uncertainty affecting project governance &
Qualitative governance assessment, scenario-based
institutional analysis, regulatory framework
evaluation &
Policy compliance, governance effectiveness,
regulatory implementation quality &
Institutional and regulatory framework design &
Governance assessed under stable institutional conditions—how regulatory frameworks behave when physical, operational, or cyber assumptions fail—central to the Texas 2021 case discussed in §3.6, remains systematically unexamined &
\cite{10.1093/polsoc/puaf006, en18061437}\\ \cline{2-8}
 
&
Eco.--Clim. &
Climate policy stringency, energy transition
pathways under climatic uncertainty, sector
coupling and hydrogen strategy governance &
Scenario-based transition planning, multi-period
expansion with endogenous carbon pricing,
partial equilibrium models for sector coupling &
Decarbonization trajectories, sector-coupling
penetration, carbon pricing levels,
electrification pathways &
Climate-aligned regulatory and investment
frameworks for energy transition &
Climate-aligned transition planning under smooth institutional implementation, absence of all three endogenous dimensions prevents detection of governance fragility when physical, operational, or cyber assumptions fail simultaneously &
\cite{en18092389, en18092205, AGHAHOSSEINI2023120401,
      vanDerZwaan2025SectorCoupling, YILMAZ2022119538,
      FLORES2024123464}\\ \cline{2-8}
 
&
Eco.--Phys. &
Regulatory uncertainty affecting DER investment,
technology cost trajectories under policy
constraints &
Risk-constrained multi-period investment model
integrating technology costs and regulatory
uncertainty for DER deployment &
Investment adequacy, DER deployment under
regulatory constraints, technology cost
recovery &
Regulatory-aware DER investment planning &
Planning-context only, bidirectional feedback through which extreme events reshape regulatory environments and through which policy constraints amplify operational fragility, both excluded, remains unrepresented &
\cite{TO2022119210}\\
 
\end{longtable}
\twocolumn
}



\small
\bibliographystyle{elsarticle-num} 
\bibliography{biblio.bib}







\end{document}